\documentclass[twocolumn]{aastex631}
\usepackage{textcomp}
\usepackage{amssymb}
\usepackage{rotating}

\newcommand{\fermi}{\textit{Fermi}}
\newcommand{\gr}{$\gamma$-ray}
\newcommand{\snr}{G65.1$+$0.6}
\newcommand{\psr}{J1954$+$2836}

\shorttitle{$\gamma$-ray Detection of the SNR~\snr}
\shortauthors{Xing et al.}

\begin{document}

\title{High-energy studies of the 3HWC J1954+286 region: likely Gamma-ray 
detection of the supernova remnant G65.1+0.6}

\correspondingauthor{Yi Xing \& Zhongxiang Wang}
\email{yixing@shao.ac.cn; wangzx20@ynu.edu.cn}

\author{Yi Xing}
\affiliation{Key Laboratory for Research in Galaxies and Cosmology,
Shanghai Astronomical Observatory, Chinese Academy of Sciences,
80 Nandan Road, Shanghai 200030, China}

\author{Dong Zheng}
\affiliation{Department of Astronomy, School of Physics and Astronomy, Yunnan University, Kunming 650091, China}

\author[0000-0003-1984-3852]{Zhongxiang Wang}
\affiliation{Department of Astronomy, School of Physics and Astronomy,
Yunnan University, Kunming 650091, China}
\affiliation{Key Laboratory for Research in Galaxies and Cosmology,
Shanghai Astronomical Observatory, Chinese Academy of Sciences, 
80 Nandan Road, Shanghai 200030, China}

\author{Xiao Zhang}
\affiliation{School of Astronomy \& Space Science, Nanjing University,
163 Xinlin Avenue, Nanjing 210023, China}
\affiliation{Key Laboratory of Modern Astronomy and Astrophysics, Nanjing University, Ministry of Education, China}

\author{Yang Chen}
\affiliation{School of Astronomy \& Space Science, Nanjing University,
163 Xinlin Avenue, Nanjing 210023, China}
\affiliation{Key Laboratory of Modern Astronomy and Astrophysics, Nanjing University, Ministry of Education, China}

\author{Guangman Xiang}
\affiliation{University of Chinese Academy of Science, Beijing 100049, China}
\affiliation{Key Laboratory for Research in Galaxies and Cosmology,
Shanghai Astronomical Observatory, Chinese Academy of Sciences, 
80 Nandan Road, Shanghai 200030, China}

\begin{abstract}
We carry out high-energy studies of the region of the Galactic TeV source 
	3HWC J1954+286, 
	whose location coincides with those of PSR~J1954+2836 and supernova 
remnant (SNR) G65.1+0.6.
	Analyzing the GeV $\gamma$-ray data obtained with the Large Area
	Telescope (LAT) onboard {\it the Fermi Gamma-ray Space Telescope},
	we are able to separate the pulsar's emission from that of the region.
Excess power-law--like emission of a $\sim 6\sigma$ significance level at the
	region is found, for which we explain as arising from 
	the SNR~G65.1+0.6. Given the low-significance detection, either a 
hadronic or 
	a leptonic model can provide a fit to the power-law spectrum. 
	Considering the properties of the pulsar and the SNR, we 
	discuss the possible origin of the TeV source, and suggest that it
	is likely the TeV halo associated with the pulsar.
\end{abstract}


\section{Introduction}
Recent very-high energy (VHE; $>$100\,GeV) surveys have shown that
our Galaxy may contain many TeV sources, as approximately 
100 such sources have thus far been detected. Notably 
there are 78 sources reported by the High Energy 
Spectroscopy System (HESS) Galactic plane survey \citep{hess18}
to be located along the latitudes of $\leq$3\arcdeg\ within
longitudes from 250\arcdeg\ to 65\arcdeg,
and there are 65 sources (a few are common detections with the HESS) 
reported by the High-Altitude Water Cherenkov (HAWC) Observatory 
in the third HAWC catalog (3HWC; \citealt{3hawc}).
While sources like the supernova remnants (SNRs) and pulsar wind nebulae 
(PWNe) are considered as primary VHE emitters, mostly due to leptonic 
processes of high-energy particles produced in them, 
a significant fraction of the reported
TeV sources can not be associated with any potential VHE 
sources \citep{hess18,3hawc}. Thus a question is raised about the nature
of the TeV sources without known counterparts: whether there are un-revealed
SNRs/PWNe or they are some other types of VHE sources. One possibility, 
recently invoked by the detections of extended TeV emissions around 
the Geminga and Monogem pulsars \citep{abe+17}, is that pulsars could commonly
produce (detectable) TeV halos \citep{lin+17}
through the inverse-Compton scattering process
of $\sim$10\,TeV electrons/positrons emitted from pulsars. Indeed, a few of the
TeV sources do appear to be positionally coincident with relatively energetic
pulsars \citep{hesspwn18,3hawc}, although it should be noted that the positions
of the sources determined from VHE observations generally  have large 
uncertainties, 
$\sim$0\fdg1.
\begin{figure}
\centering
\includegraphics[width=0.47\textwidth]{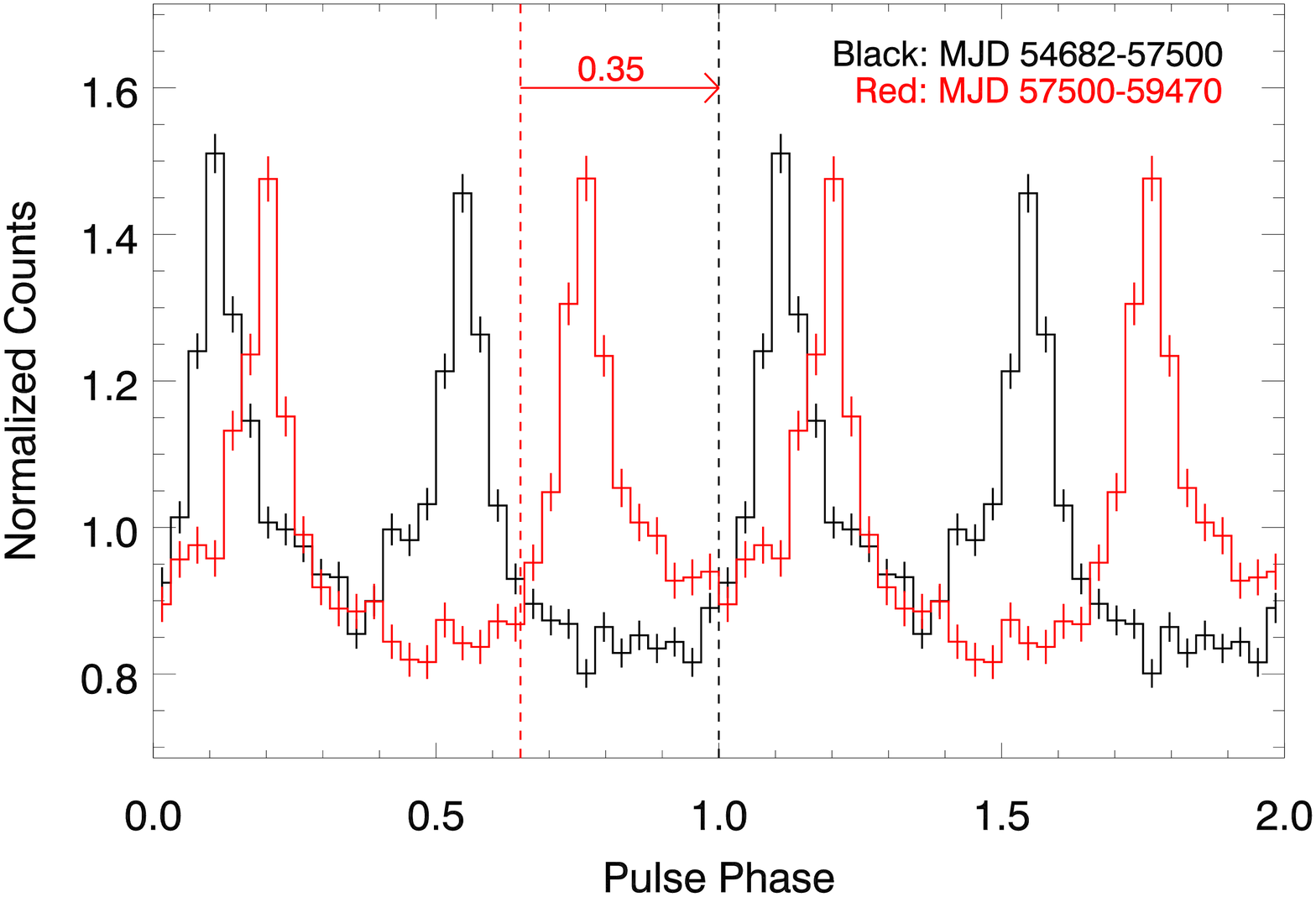}
\caption{Phase-connected pulse profiles of \psr\ during MJD 54682--57500 
	(black) 
	and MJD 57500--59470 (red). The zero phases of each profiles are 
	marked as dashed lines. Two spin cycles are shown for clarity.}
\label{fig:p2}
\end{figure}

For the purpose of investigating the nature of the TeV sources not in obvious
associations,
we took advantage of the available different high-energy data, in particular 
the all-sky survey ones from the Large Area Telescope (LAT) onboard 
{\it the Fermi Gamma-ray Space Telescope (Fermi)}, and
have carried out systematic studies of the TeV sources by using 
a multi-wavelength study approach. 
In this paper, we report our study of the region of
3HWC J1954$+$286.

For this VHE source, initially 
excess TeV emission with a 4.3$\sigma$ significance was found
with Milagro at the position of the \fermi\ LAT source 0FL~J1954.4+2838
\citep{abd+09_Milagro}.
A TeV source named 2HWC J1955$+$285 was listed in the second HAWC catalog 
in this region with a Test Statistic (TS) value of 25.4 \citep{2hawc}, 
and this source's TS value was increased to be 48.3 in 3HWC (with an 1$\sigma$
positional uncertainty of 0\fdg14). Furthermore, a source was  
detected to have the maximum emission energy of 0.42\,PeV
with the Kilometer Square Array (KM2A) of the Large High Altitude Air Shower 
Observatory (LHAASO; named LHAASO J1956+2845; \citealt{lhaaso_pev_2021}), 
although its position is 0\fdg33 away from the 3HWC position.
Therefore the nature of this TeV source, which even has a possible
PeV counterpart, is extremely interesting.

As pointed out by \citet{2hawc}, a young pulsar, PSR~\psr, is known to be 
located 
within the 1$\sigma$ error circle of 3HWC~J1954$+$286. Its spin-down luminosity 
$\dot{E}\simeq 1.0\times 10^{36}$\,erg\,s$^{-1}$, and it was detected in
$\gamma$ rays with \fermi\ LAT \citep{par+10}, having a \gr\ 
luminosity
of $\sim 4.1\times 10^{34}$\,erg\,s$^{-1}$ (where a dispersion-measurement
distance of 1.96\,kpc is used). 
It can be noted that there is a SNR, \snr, also in this source region
\citep{abd+09_Milagro}.
This SNR was discovered in the radio band with low surface 
brightness \citep{lcp90,kot+06}. It has a large SNR-like shell with a size 
of $\sim$90$^\prime$$\times$50$^\prime$ \citep{lcp90,tl06}. 
A distance of 9.2\,kpc was derived for this SNR based on the HI observations, 
and a Sedov age of 4--14$\times$10$^{4}$ yr was estimated \citep{tl06}. 
\begin{figure}
\centering
\includegraphics[width=0.47\textwidth]{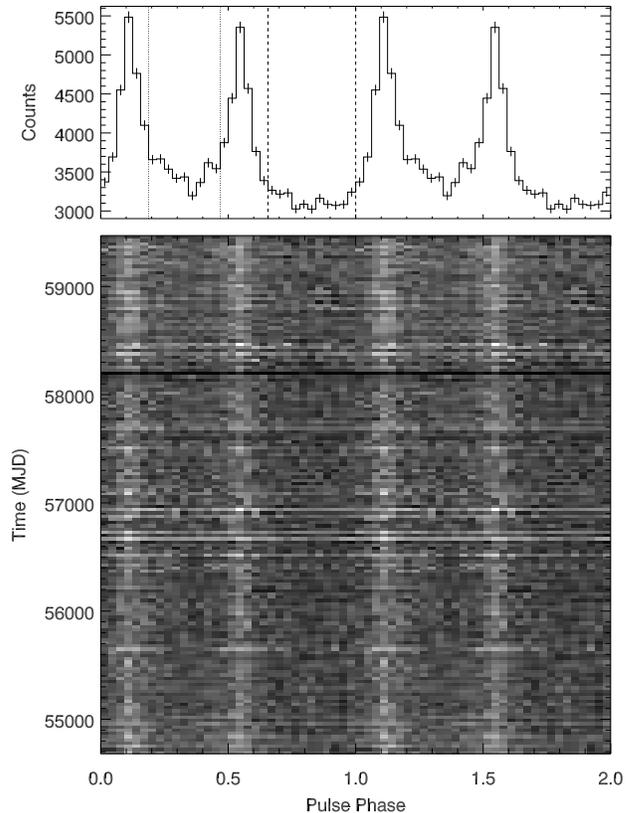}
	\caption{Phase-connected pulse profile ({\it top}) and two-dimensional 
	phaseogram ({\it bottom}) in 32 phase 
	bins obtained for PSR~\psr\ during the whole LAT data time period. 
	The dotted and dashed lines mark the bridge and offpulse phase 
	ranges respectively. Two spin cycles are shown for clarity.}
\label{fig:prof}
\end{figure}

Below we first describe our analysis of the \fermi\ LAT data in 
Section~\ref{sec:obs} for the region, in which we were able to remove 
the emission of the pulsar. We also analyzed archival X-ray data covering
the region. The results from the analyses are presented 
and discussed in Section~\ref{sec:dis}.


\begin{figure*}
\centering
\includegraphics[width=0.32\textwidth]{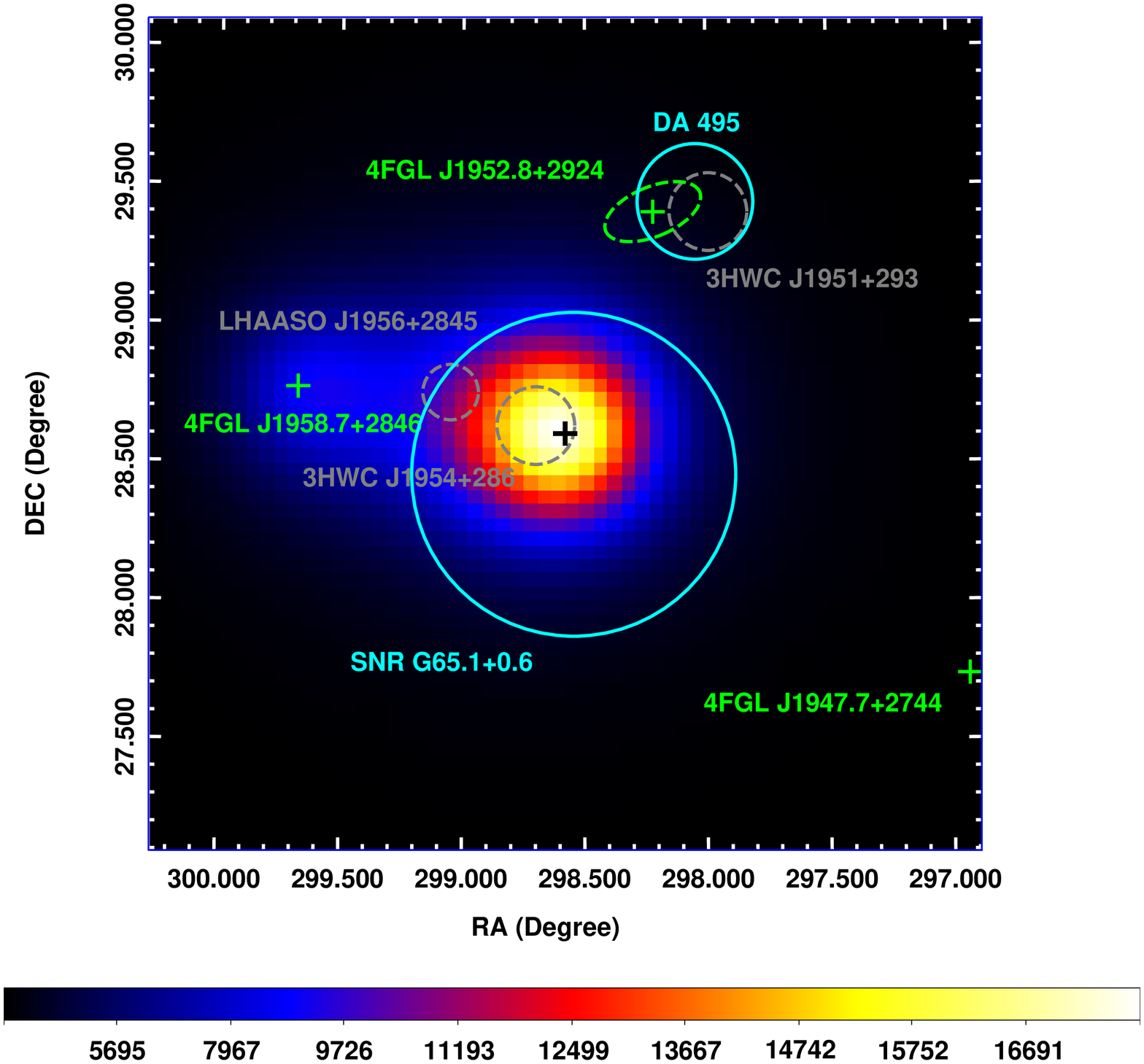}
\includegraphics[width=0.32\textwidth]{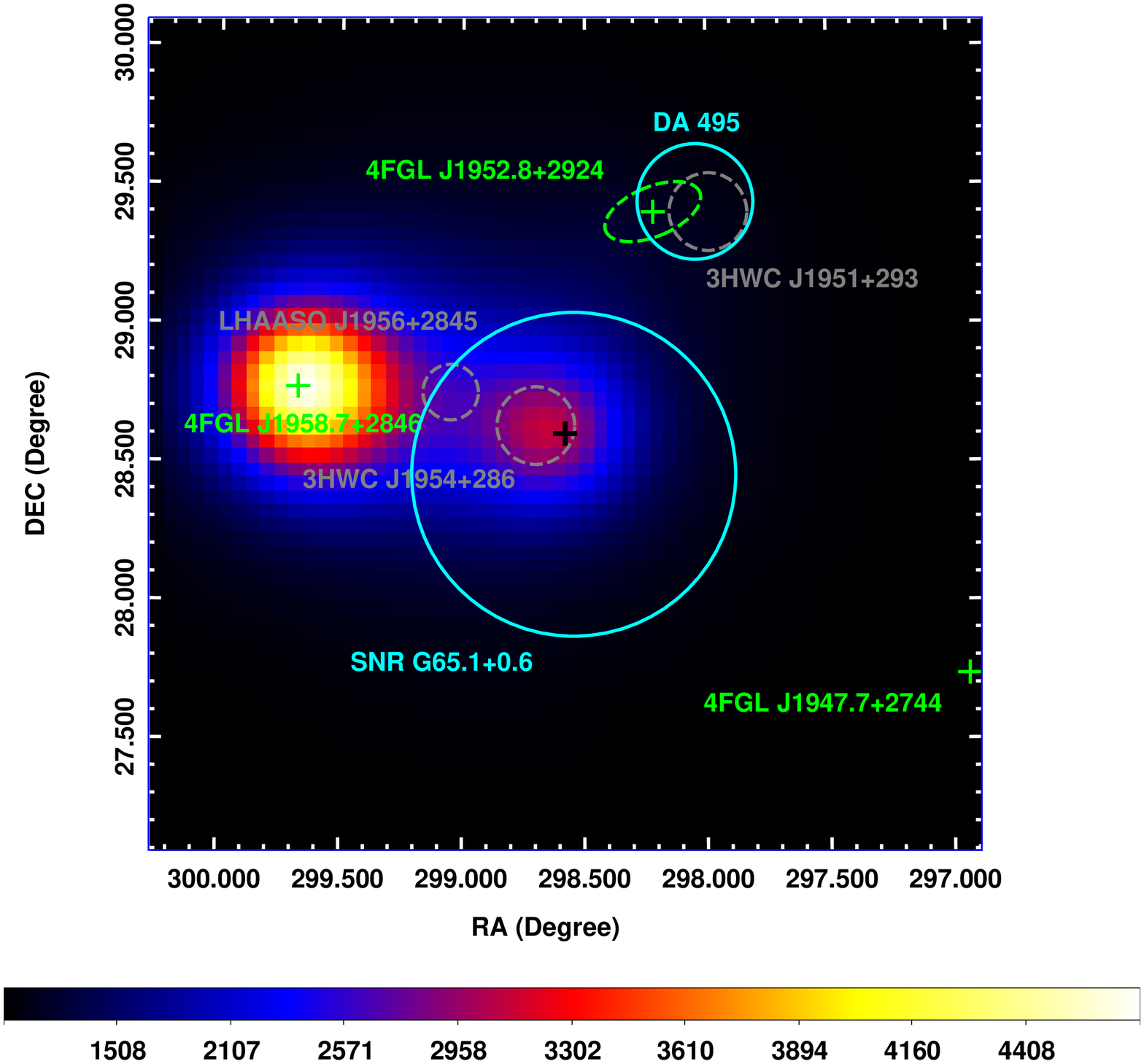}
\includegraphics[width=0.32\textwidth]{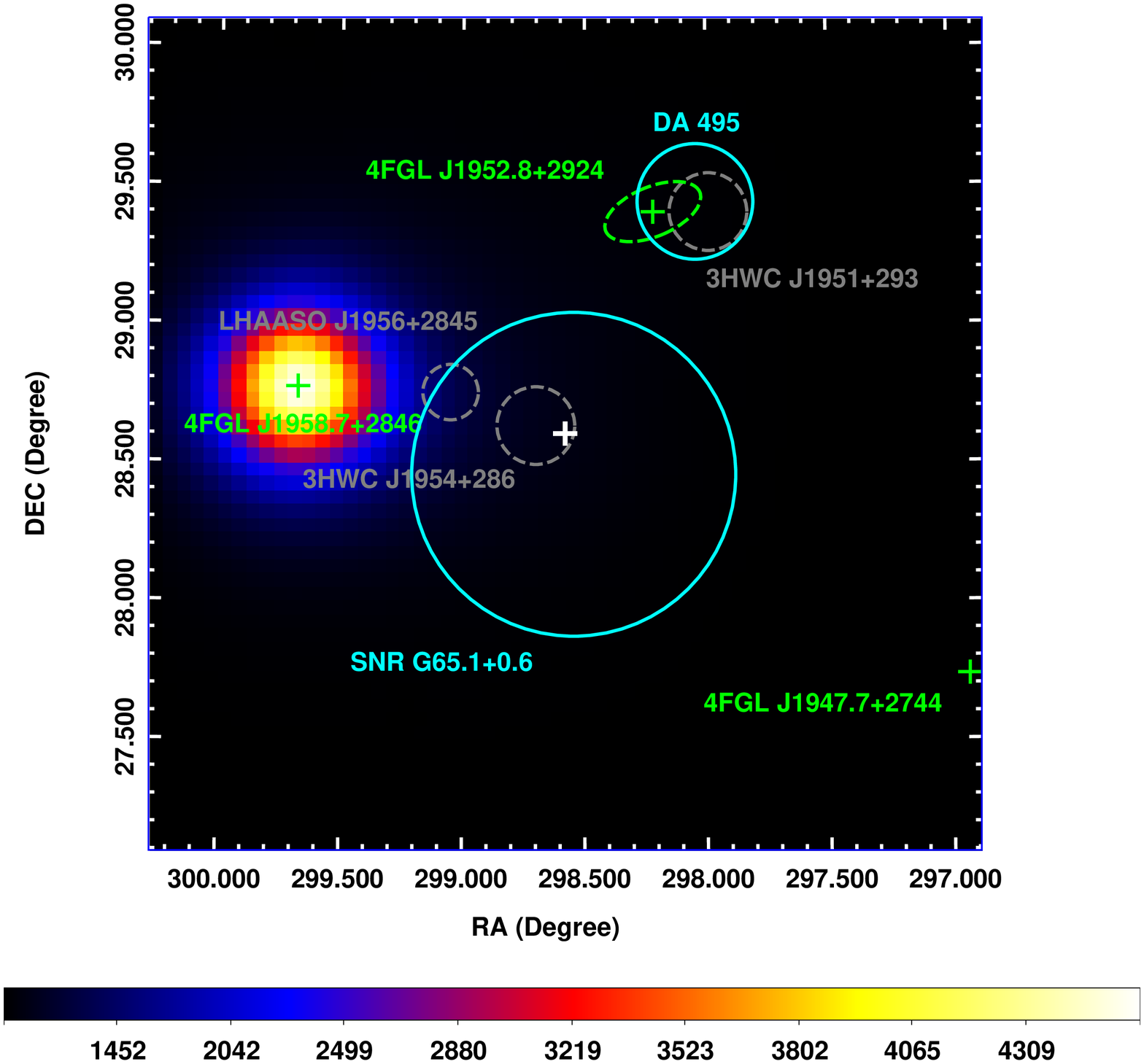}
	\caption{TS maps (in 0.1--500\,GeV) for the region of PSR~\psr\ 
	during the pulsar's onpulse ({\it left}), bridge ({\it middle}),
	and offpulse ({\it right}) phase ranges.
All sources in the source model are kept.  For each TS maps, the image scale 
	is 0.05 degree pixel$^{-1}$ and the 
	color bar indicates the TS value range. 
	The black (or white) and green pluses mark the positions of 
	PSR~\psr\ and 
	catalog sources in 4FGL-DR2 respectively, while the green 
	dashed line indicates the 2$\sigma$ error ellipse for the catalog 
	source in the DA 495 region. 
	The cyan circles represent 
	the $\sim$70\,arcmin extension of the radio emission from 
	the SNR~\snr\ \citep{tl06} and DA~495 \citep{kot+08}. 
	The gray dashed lines mark the 1$\sigma$ error circles of two HAWC 
	sources \citep{3hawc} and the reported LHAASO KM2A source.
	}
\label{fig:tsmap-psr}
\end{figure*}

\section{Data Analysis}
\label{sec:obs}

\subsection{LAT data analysis}

\subsubsection{LAT data and source model}
\label{sec:ld}

We used the \gr\ data collected with \fermi\ LAT \citep{atw+09}
since 2008 August. 
The region of interest (RoI) was set centered 
at the position of PSR~\psr\ with 
a size of $20\arcdeg\times 20\arcdeg$.
The events in this RoI in the energy range of from 50 MeV to 500 GeV 
over the time period from 2008-08-04 15:43:36 (UTC) to 
2021-09-13 22:13:15 (UTC) were selected. The latest 
\fermi\ Pass 8 database was used.
Following the recommendations of the LAT team\footnote{\footnotesize http://fermi.gsfc.nasa.gov/ssc/data/analysis/scitools/}, 
the events with quality flags of `bad' and 
zenith angles larger than 90 degrees were excluded.

A source model was constructed on the basis of the recently released 
\fermi\ LAT 10-year source catalog (4FGL-DR2; \citealt{4fgl20,bal+20}).
The sources in 4FGL-DR2 within a 20 degree radius 
circular region from PSR~\psr\ were included in the source model. 
Their spectral forms provided in the catalog were used.
The background Galactic and extragalactic diffuse 
spectral models (gll\_iem\_v07.fits and iso\_P8R3\_SOURCE\_V3\_v1.txt
respectively) were also included in the source model.

\subsubsection{Timing analysis of PSR~\psr}
\label{sec:ta}

PSR~\psr\ is bright in the LAT $\gamma$-ray band, and so we first worked
to separate its emission from other sources in the region through timing
analysis. 
Following \citet{par+10}, we selected photons within a
1\arcdeg\ radius circular region centered at the pulsar in order to
construct its pulse profile.  We tested to fold them with the 
ephemeris given in the second \fermi\ LAT catalog of \gr\ pulsars 
(\citealt{abd+13}; see also Table~\ref{tab:search_result}, with
the first set of the parameters from the online material of 
\citealt{abd+13}). 
No clear pulse profile over the 13-yr--long time period could be obtained.

We changed to assign pulse phases to the photons during  
MJD~54682--55800 (the same time period as that in \citealt{abd+13}) 
according to the known ephemeris, using the \fermi\ TEMPO2 
plugin \citep{edw2006,hob2006}.  An empirical Fourier template profile was
extracted. Using this template, we were able to generate 
the time-of-arrivals (TOAs) 
for each sets of $\sim$200-day data over the whole LAT data time period. 
The template and TOAs were obtained using the maximum likelihood method 
described in \citet{ray2011}. 
We fitted the TOAs with TEMPO2. 
The known ephemeris has three timing parameters: $f$, $f1$, and $f2$
(Table~\ref{tab:search_result}).
We updated the ephemeris by adding high-order frequency derivatives, while we
tried including as many 200-day datasets as possible. However 
an updated ephemeris (its $f$, $f1$, and $f2$ are given in 
Table~\ref{tab:search_result}) only describing the TOAs before 
MJD $\sim$57500 was obtained.

For the data after MJD $\sim$57500, we were able to find 
a template profile during MJD 58300--59470 (where the latter date
is the end time of the whole data), by searching around the pulsar catalog 
ephemeris values.  Same as the above by adding high-order frequency 
derivatives,
we obtained an ephemeris that can describe the TOAs after MJD 57500 
(the main parameters are given in Table~\ref{tab:search_result}). 
\begin{figure*}
\centering
\includegraphics[width=0.3\textwidth]{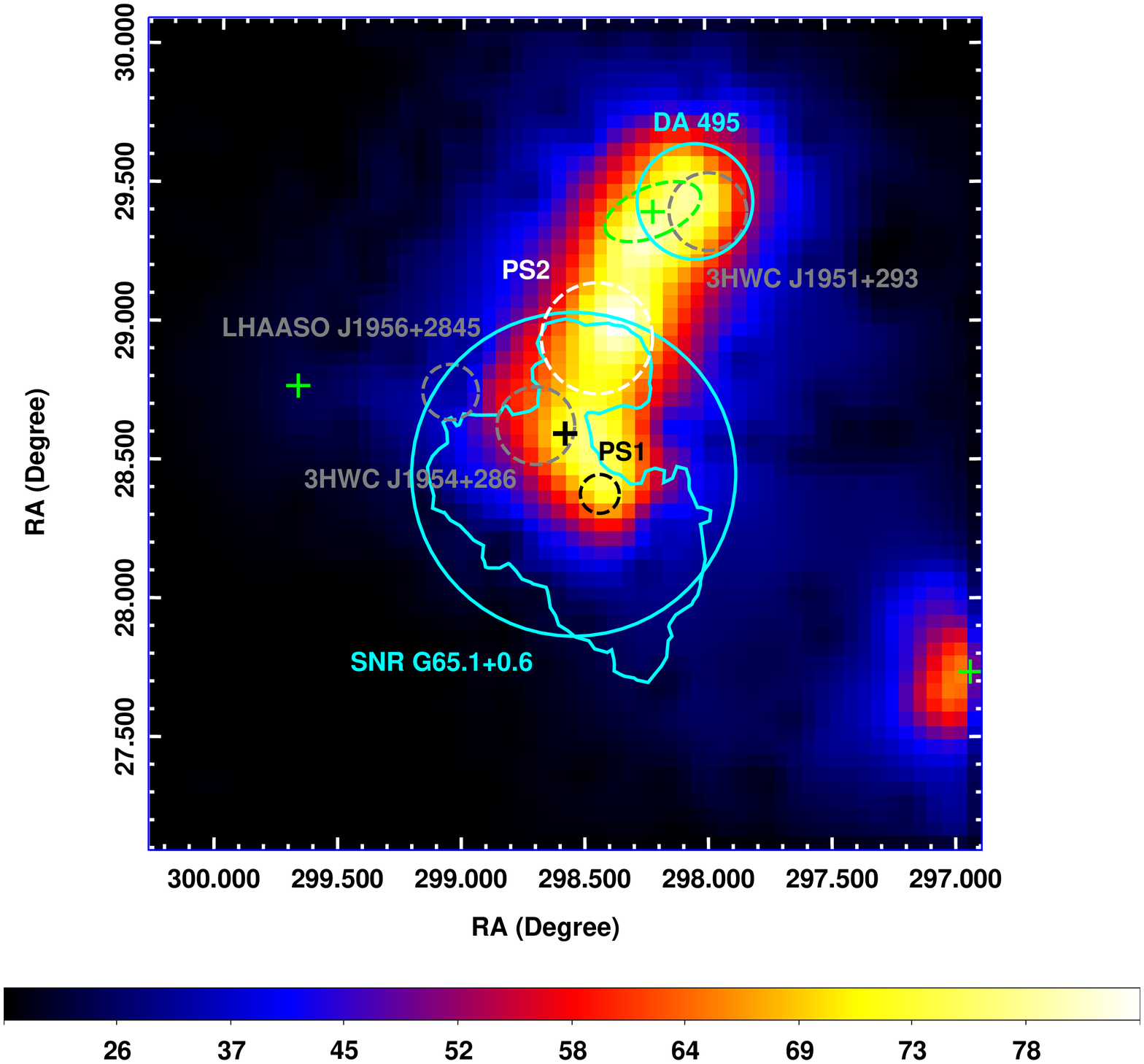}
\includegraphics[width=0.3\textwidth]{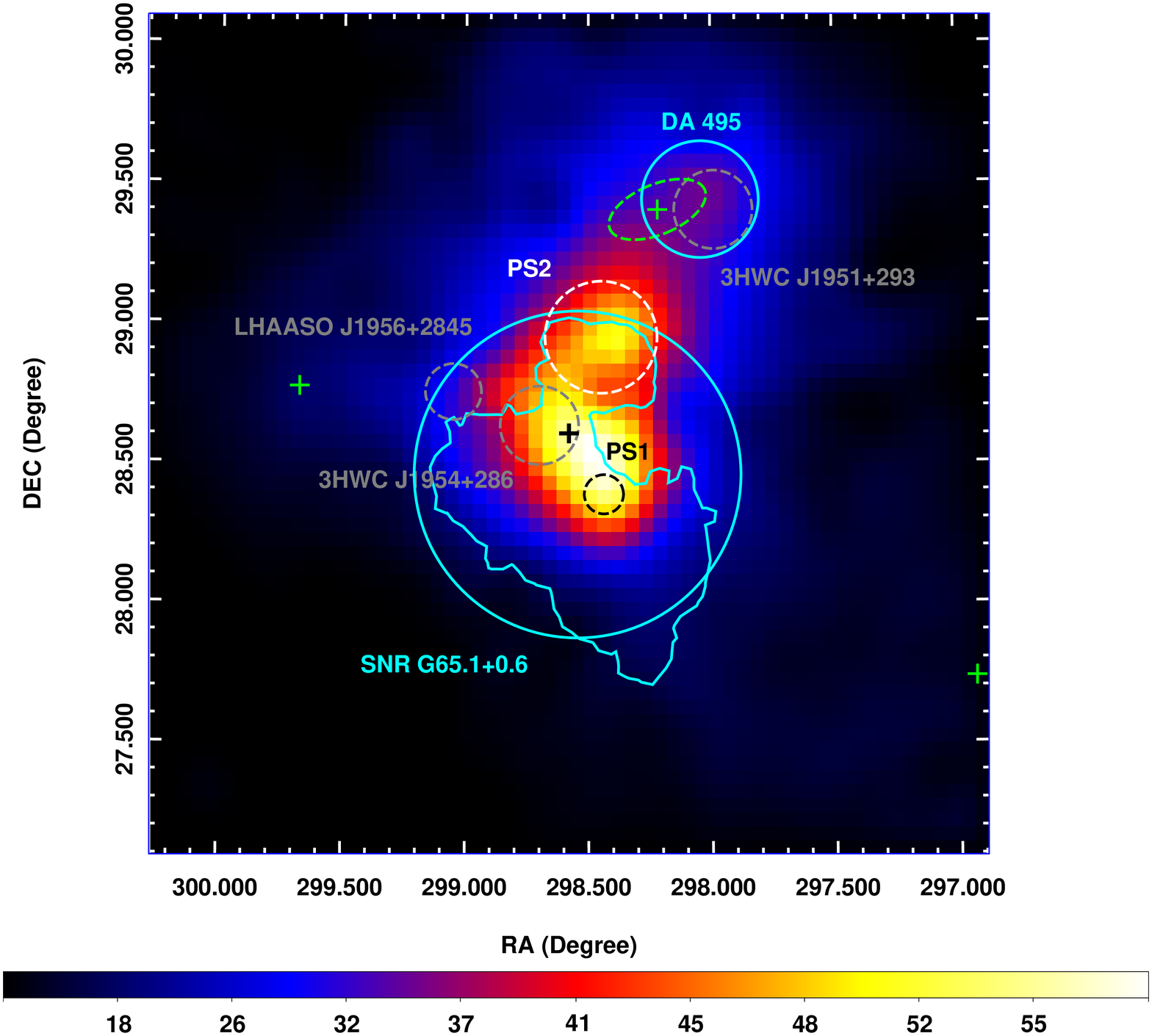}
\includegraphics[width=0.3\textwidth]{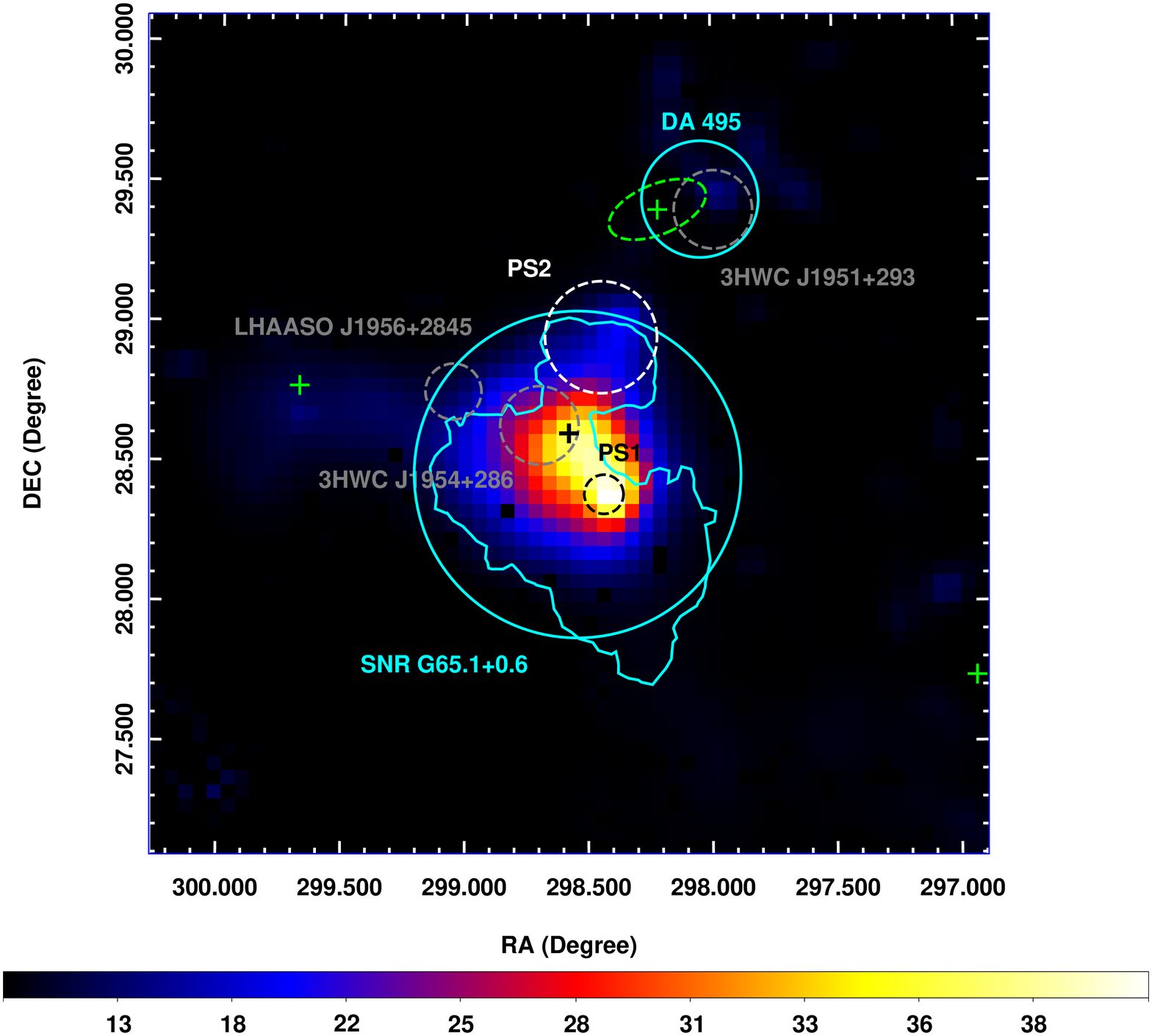}
\caption{TS maps for the region of PSR~\psr\ during the pulsar's offpulse 
	phase range (with the source model not including the pulsar). 
	The {\it left} and {\it middle} panels respectively show
	the region in 0.1--500 GeV band with only the bright catalog 
	source 
	4FGL~J1958.7+2846 (cf., Figure~\ref{fig:tsmap-psr}) removed and 
	all catalog sources removed, and
	the {\it right} panel shows the region in 0.5--500 GeV band with 
	all catalog sources removed.
	The cyan contour is an approximate radio shape of the SNR~\snr\ 
	(obtained from the SNRcat: http://snrcat.physics.umanitoba.ca/SNRrecord.php?id=G065.1p00.6), and the black and white dashed lines mark 
	the 2$\sigma$ error circles centered at the best-fit positions 
	for PS1 and PS2 (cf., Section~\ref{sec:off}). Other marks are the same
	as those in Figure~\ref{fig:tsmap-psr}.
	}
\label{fig:tsmap}
\end{figure*}

We folded the LAT photons before and after MJD~57500 according to the 
two ephemerides respectively. The folded pulse profiles are shown 
in Figure~\ref{fig:p2}. We cross-correlated the two profiles in the Fourier 
domain and obtained a phase shift of $\simeq$0.35 (Figure~\ref{fig:p2}). 
Applying the phase shift to the second part of the photons, the pulse profile 
over the whole data period was obtained. In Figure~\ref{fig:prof},
the pulse profile and two-dimensional phaseogram in 32 phase bins 
are plotted. Based on the profile, we defined phase 0.0--0.19 and 
0.47--0.66 as the onpulse phase ranges, phase 0.19--0.47 as a bridge phase 
range, and phase 0.66--1.0 as the offpulse phase range.


\subsubsection{Likelihood analysis for the onpulse and bridge data}
\label{sec:la}

We performed standard binned likelihood analysis
to the 0.1--500 GeV LAT data during the onpulse and bridge phase ranges
of the pulsar determined above.
The RoI and the source model, set in Section~\ref{sec:ld},
were used.
The sources in our source model within 5 degrees from 
the pulsar were set to have free spectral parameters; for the other sources, 
their spectral parameters were fixed at the values given in the catalog.
The background normalizations were set as the free parameters.

For the emission at the pulsar's position in the two phase ranges, 
which is presumably dominated by that from the pulsar 
(Figure~\ref{fig:tsmap-psr}), we used a sub-exponentially cutoff 
power-law model,
$dN/dE = N_{0}E^{-\Gamma}\exp[-(E/E_{c})^{b}]$,
where $\Gamma$ and $E_{c}$ are the photon index and cutoff energy respectively,
and $b$ is a measure of the shape of the exponential cutoff. We fixed $b$ 
to be $2/3$, which is a characteristic value used for the \gr\ pulsars 
in 4FGL-DR2. We obtained $\Gamma=1.42\pm0.01$ and $E_c=1.60\pm0.03$
($\Gamma=1.29\pm0.06$ and $E_c=0.96\pm0.07$) for the onpulse (bridge) phase
range. The values are close to those given in 4FGL-DR2 ($\Gamma=1.43\pm0.06$,
$E_c=1.4\pm0.1$\,GeV for the whole data).
These results, as well as the corresponding Test Statistic (TS) values,
are provided in Table~\ref{tab:likelihood}. 

In order to show the appearances of the pulsar and other detected 
sources during the different phase ranges of the pulsar,
we calculated the 0.1--500 GeV TS maps of
a $3\arcdeg\times 3\arcdeg$ region centered at the pulsar 
(Figure~\ref{fig:tsmap-psr}).
All catalog sources in the source model were kept in the TS maps. The pulsar
was dominantly bright during the onpulse phase range, and in the bridge and
offpulse phase ranges, a nearby source, 4FGL~J1958.7+2846, appeared brighter.

\begin{figure}
\centering
\includegraphics[width=0.47\textwidth]{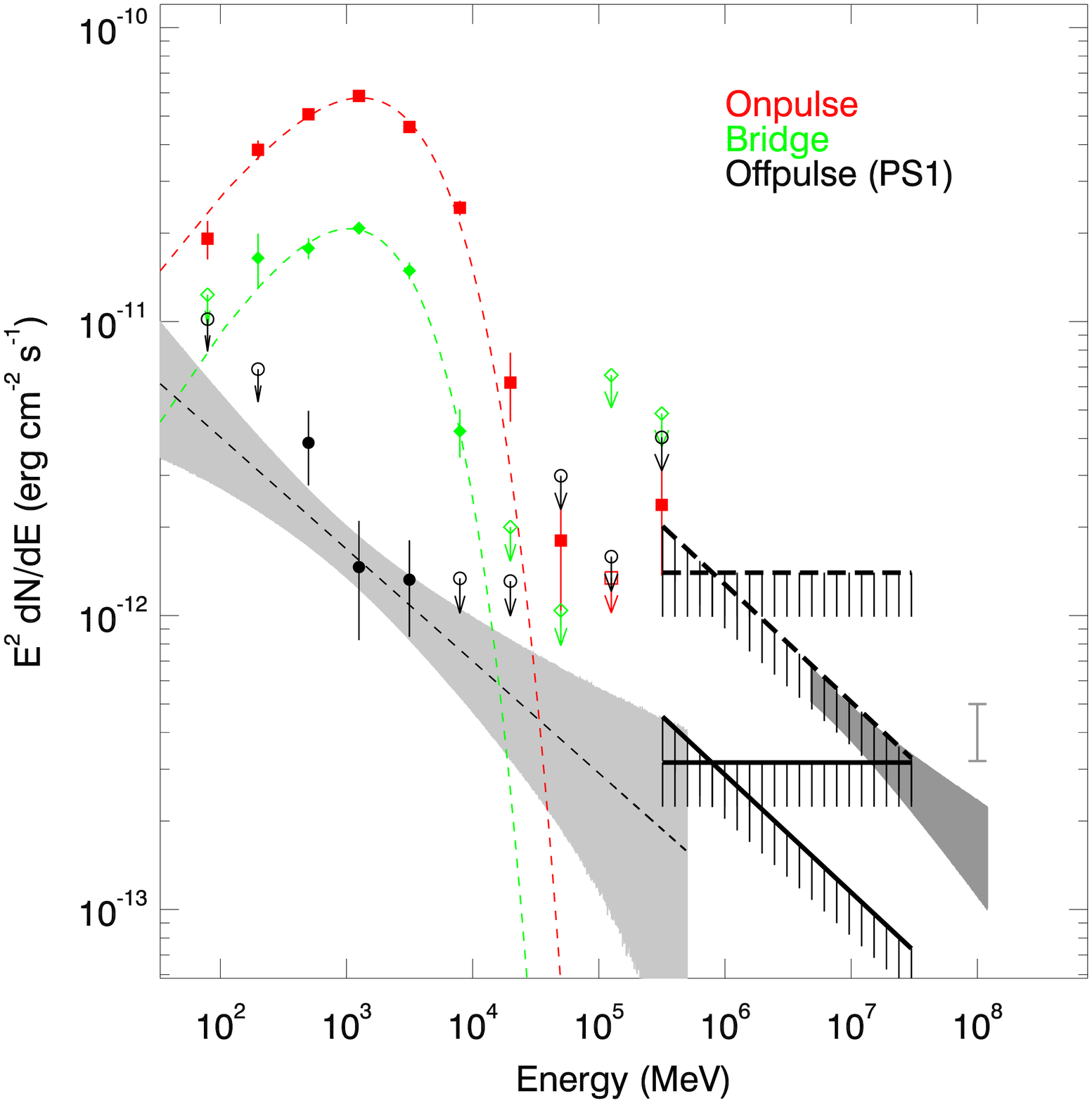}
\caption{\gr\ spectra and corresponding 0.1--500 GeV spectral models 
of PSR~\psr\ during the onpulse and bridge phase ranges (red and green 
respectively), and of PS1 during the offpulse phase range (black and light
	gray area).
The dark gray area is the HAWC spectrum for 3HWC J1954$+$286 
(\citealt{3hawc}), and
the gray bar is the flux measurement for LHAASO J1956$+$2845 at the energy 
of 100 TeV (\citealt{lhaaso_pev_2021}). 
The thick solid lines and long-dashed 
lines mark the upper limits obtained with the VERITAS from the point-source 
search and extended-source search, respectively (the spectral indices of 
2 and 2.4 were considered; see \citealt{abe+18}).
}
\label{fig:spectra}
\end{figure}

\subsubsection{Likelihood analysis for the offpulse data}
\label{sec:off}

For the offpulse phase range, if we still assumed the sub-exponentially 
cutoff model for the 0.1--500 GeV \gr\ emission at the position of PSR~\psr, the 
cutoff could not be significantly determined. If we instead used a simple 
power law $dN/dE = N_{0}E^{-\Gamma}$, we would obtain
$\Gamma=2.6\pm0.1$, which is significantly larger than those in
the onpulse and bridge phase ranges.

In order to examine the region, we calculated 0.1--500 GeV TS maps of
a $3\arcdeg\times 3\arcdeg$ region centered at the pulsar. The left 
and middle panels of Figure~\ref{fig:tsmap} respectively show the region 
with all catalog sources kept (except 4FGL J1958.7$+$2846)
and removed; the nearby source 4FGL J1958.7$+$2846
(cf., Figure~\ref{fig:tsmap-psr})
was removed in the left panel such that the other sources in 
the region can be relatively well seen. 

The TS map shown in the middle panel of Figure~\ref{fig:tsmap} indicates 
that the residual \gr\ emission is obviously not centered at the pulsar; 
moreover,
there appear to likely have two point sources. 
We ran \textit{gtfindsrc} in {\tt Fermitools} to determine their positions.
The obtained best-fit positions are (R.A., Decl.) $=$ (298\fdg44, 28\fdg38) and
(R.A., Decl.) $=$ (298\fdg4, 28\fdg9) for point source 1 (PS1) and 2 (PS2), 
respectively, with the 2$\sigma$ nominal uncertainties of 0\fdg07 and 0\fdg2
(see Figure~\ref{fig:tsmap}). The position for PS1 seems to be off from
the highest-TS location. 
However as we checked TS maps at different higher energy bands such as 
0.3--500, 0.5--500, and 1--500\,GeV, we found that the PS1 position is 
right at the locations with the highest TS values in them.
The 0.5--500 GeV TS map is shown as an example in the right panel 
of Figure~\ref{fig:tsmap}.
The localization result obtained with 0.1--500 GeV data is thus 
likely correct, as the large Point Spread Function (PSF) of LAT 
(possibly resulting in contamination from nearby sources) 
as well as the strong Galactic background in the low energy range of 
$<$300\,MeV can cause the off-peak appearance.
We also noted that in the 0.5--500 GeV TS map,
PS2 is nearly invisible, indicating that it had very soft emission and 
likely is a separate source not in association with PS1.

As the location of PS1 is away from the pulsar but within the SNR~\snr\ 
(Figure~\ref{fig:tsmap}),
we considered it as a possible counterpart to the SNR and re-performed 
the likelihood analysis to the LAT data during the offpulse phase range, 
in which we removed PSR~\psr\ from the source model and included PS1.
A simple power law was assumed for the source, 
and we obtained $\Gamma=2.4\pm0.1$ (the details are given in 
Table~\ref{tab:likelihood}). We noted that the radio flux density contour 
of the SNR is much extended and contains PS2, and so we tested to use 
the radio contour as a template. The likelihood analysis was conducted
by including this template in the source model. The likelihood value 
$L_{\rm ext}$ was compared to that $L_{\rm 2PS}$ when PS1 and PS2 were
considered as two point sources or that $L_{\rm PS1}$ when only PS1 was 
considered.
The resulting results were $-2\times log(L_{\rm ext}/L_{\rm 2PS})\sim$23 or
$-2\times log(L_{\rm ext}/L_{\rm PS1})\sim$16. As either value is significantly
large, suggesting that no extended \gr\ emission was preferred over the 
individual point sources.
\begin{figure*}
\centering
\includegraphics[scale=0.42,angle=0]{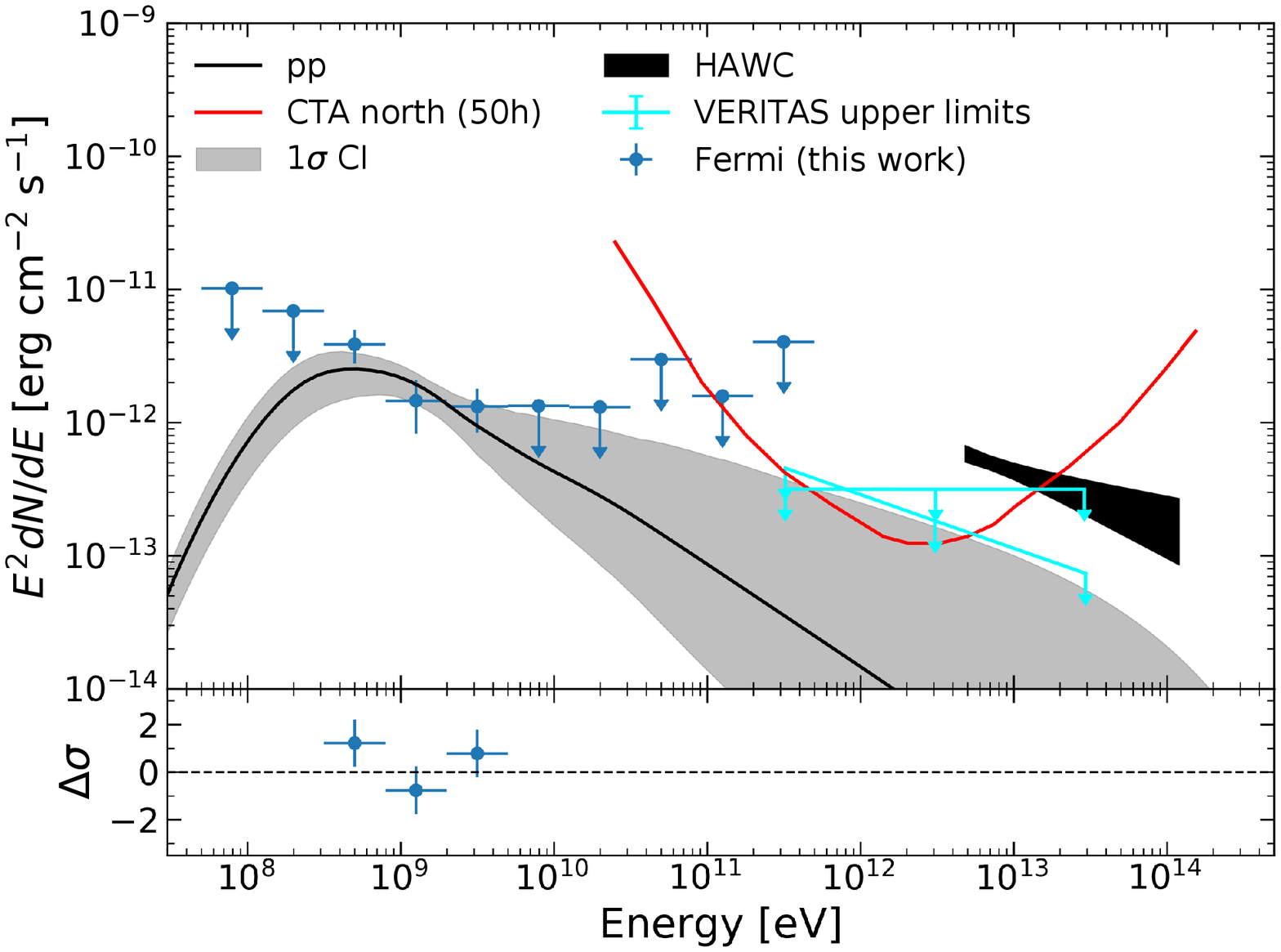}
\includegraphics[scale=0.42,angle=0]{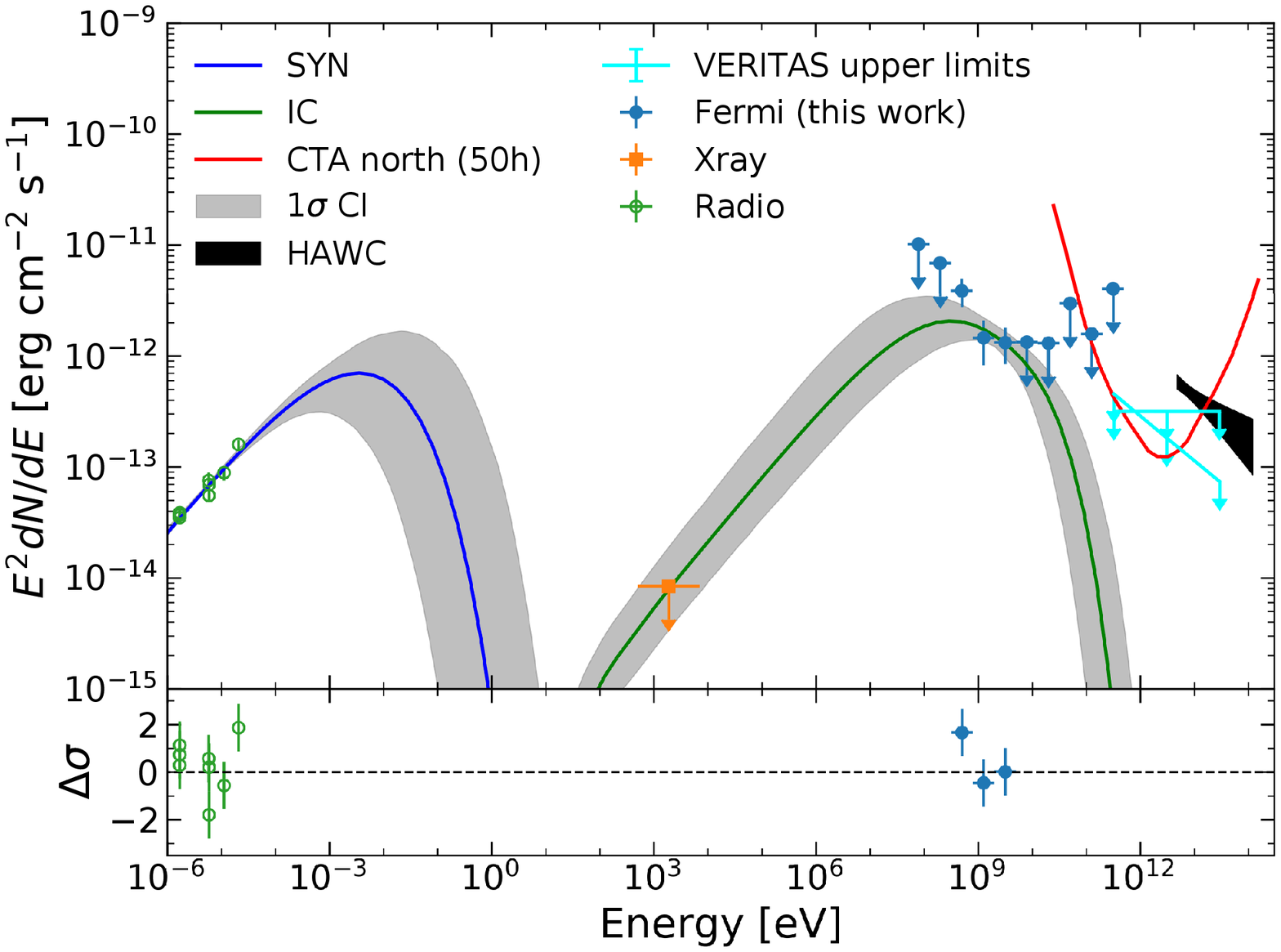}
	\caption{Fitting to the SED of SNR G65.1+0.6. The gray band 
	corresponds to the $1\sigma$ confidence interval (CI). 
	The VERITAS upper
	limits from the point-source search for the region and the
	fluxes of 3HWC J1954+286 are shown as the cyan lines and dark black 
	area respectively. The sensitivity 
	of CTA north \citep{cta2019.book} is displayed as a red solid line.
	{\it Left} panel: hadronic model. {\it Right} panel: leptonic model,
	in which the green data points are radio flux measurements
	from \citet{lcp90,kot+06,tl06,gao+11} and the yellow data
	point is the X-ray flux
	upper limit derived from the {\it Chandra} observation in this work.
	}
\label{fig:sed}
\end{figure*}


\subsection{Spectral and variability analysis}
\label{sec:sva}

We extracted the \gr\ spectra of PSR~\psr\ during the onpulse and bridge
phase ranges and that of PS1 during the offpulse phase range.
The maximum likelihood analysis of the LAT data in 10 evenly divided energy 
bands in logarithm from 0.05 to 500\,GeV was conducted.
In this analysis, the spectral normalizations of the sources within 5 degrees 
from each target were set as free parameters, while all the other parameters 
of the sources were fixed at the values obtained from the maximum 
likelihood analysis in the above section. 
For the results, we kept the spectral data points when the fluxes are 
$>$2 times larger than the uncertainties and derived 95\% flux upper limits 
otherwise. The obtained spectra are shown in Figure~\ref{fig:spectra}, and
the values are given in Table~\ref{tab:spectra}.

In order to fully study the \gr\ emission properties of PS1, we also searched 
for its long-term variability. We calculated the variability index TS$_{var}$ 
with 80 time bins (each bin was constructed from 60-day data) in the energy 
range of 0.1--500 GeV, following the procedure introduced in \citet{nol+12}. 
If its fluxes were consistent with a constant, TS$_{var}$ would be distributed 
as a $\chi^2$ function with 79 degrees of freedom.
The computed TS$_{var}$ value was 96.9, lower than the threshold value of
111.1 (at a 99\% confidence level) for identifying a variable source.
Therefore PS1 had no significant long-term variability.

\subsection{{\it Chandra} X-ray data analysis}
{\it Chandra} X-ray telescope observed PSR J1954+2836 on 2011 December
26 (Obs. ID 12148) with an effective exposure time of 9.84\,ks. 
The Advanced CCD Imaging Spectrometer array was used in the observation
to image the source region. We processed the data with the software CIAO v4.13 
and the {\it Chandra} Calibration Database (CALDB v4.9.5). Task 
{\tt wavdetect} was used for detecting sources in the field, but 
no sources were found at the pulsar's position or in the nearby regions. 
The upper limit on the
count rate at the position is 4.56$\times 10^{-4}$\,cts\,s$^{-1}$.
We estimated an unabsorbed model flux based on the count-rate upper limit.
Assuming a power law with a photon index of 1.5 and a hydrogen column density
of $N_{\rm H}\sim 1.1\times 10^{22}$\,cm$^{-2}$ towards the source position
\citep{nh16}, the flux upper limit
of $\sim 8.9\times 10^{-15}$\,erg\,s$^{-1}$\,cm$^{-2}$ 
in 0.5--7.0\,keV is given by {\it Chandra} tool {\tt PIMMS}.

\section{Results and discussion}
\label{sec:dis}

To investigate the nature of the VHE source 3HWC~J1954+286, we analyzed 
13 years of Fermi LAT Pass 8 data for the source's region. 
By determining the onpulse and bridge phase ranges of PSR~\psr\ and
using the different phase-range LAT data accordingly, 
we were able to separate the pulsar's emission from that of the region. The 
pulsar's emission in the onpulse and bridge phase ranges can be well fitted
with a typical spectral shape of pulsars, a sub-exponentially cutoff power law
\citep{abd+13,4fgl20}. After removing the pulsar's emission component 
(i.e., by using the data only during the offpulse phase range),
excess emission has been found in the region.  There are likely two point 
sources, and their determined positions are significantly away
from that of the pulsar. 
Also from the likelihood analysis of the data in 0.1--500 GeV band,
we found that the emission of each of the two sources is consistent with
being described with a power law
(see Table~\ref{tab:likelihood}). Given the positional
differences and spectral types, the two sources very likely 
are not emission from the pulsar in the offpulse phase range.

For the HAWC TeV source 3HWC~J1954+286, both PS1 and PS2 appear away
from it, having 2.3$\sigma$ and 1.9$\sigma$ positional differences 
respectively. As PS2 was faint (Table~\ref{tab:likelihood}) and limited
property information was extracted for it, below we mainly discuss the
more significantly detected PS1 ($\sim$6$\sigma$ detection significance). 
We have obtained its spectrum.
The extension of the spectrum according to its power-law index would
predict approximately an order of magnitude lower TeV fluxes than those 
detected by HAWC (Figure~\ref{fig:spectra}). Based on these facts, the excess 
emission is not likely the GeV counterpart to the TeV source either.
We note that the Very Energetic Radiation Imaging Telescope Array
System (VERITAS) also observed the region \citep{abe+18}, and the flux 
upper limits derived from the observation for a point source have reached 
the upper bound of the extension of the PS1's spectrum 
(Figure~\ref{fig:spectra}), which potentially provides a constraint
on any TeV emission of PS1. An additional note is that the HAWC source 
would be extended, such that the VERITAS upper limits become 
consistent with the HAWC flux measurements (Figure~\ref{fig:spectra};
see discussion in \citealt{abe+18}).

One possibility is if the excess emission could be the PWN associated with
the pulsar. However, if the excess emission is the PWN,
we would have detected it in X-rays. According to theoretical calculations
as well as observational evidence given in \citet{zzf18}, a pulsar 
like PSR~\psr\ would have an X-ray--to--\gr\ luminosity ratio of 
$\gtrsim5\times 10^{-3}$.
The source PS1 has a \gr\ luminosity of
$L_{\gamma}\simeq 4.5\times 10^{33} (d/1.96\,{\rm kpc})^2$\,erg\,s$^{-1}$,
where the pulsar's dispersion-measurement distance is assumed.
This \gr\ luminosity would predict an X-ray luminosity 
of $\gtrsim 2\times 10^{31}$\,erg\,s$^{-1}$, 
or $\gtrsim 4\times 10^{-14}$\,erg\,s$^{-1}$\,cm$^{-2}$ (at 1.96\,kpc).
Comparing this value to the flux upper limit set by the {\it Chandra}
observation, we would see some evidence for an X-ray source at the pulsar's 
location in the X-ray image. Therefore the non-detection
of any X-ray sources in the sensitive {\it Chandra} observation does not 
support it as the PWN.
In addition, the power-law--like spectrum of PS1 also does not support 
the PWN origin since the PWNe generally have a \gr\ spectrum similar to 
those of pulsars \citep{ack+11}, although the \gr\ efficiency 
$L_{\gamma}/\dot{E}\simeq 4.5\times 10^{-3}$ would be consistent with
the range of \gr\ efficiency values of the known PWNe \citep{ack+11}.

The possibility remaining to be considered is that the excess emission PS1 
would 
be the \gr\ counterpart to the SNR~\snr, as their positions are coincident with
each other. Below in Section~\ref{sec:snr}, we discuss possible scenarios to 
explain the excess \gr\ emission as arising from the SNR. In the end 
the section~\ref{sec:tev}, we discuss the speculation if
the TeV source could be related to the SNR or the pulsar in the region
given their properties. 

\subsection{Emission scenarios for the SNR~\snr}
\label{sec:snr}

At the SNR's distance of 9.2 kpc, the 0.1--500 GeV luminosity 
was $\simeq 1\times$10$^{35}$ erg s$^{-1}$, which is in the range of 
middle-aged SNRs ($\sim$10$^{35}$ $-$ 10$^{36}$ erg s$^{-1}$; 
\citealt{acero16}).
By interacting with dense clouds, middle-aged SNRs have 
\gr\ emission dominated by that arising from the hadronic 
process.
They may have flux peaks at $\sim$1 GeV (e.g. W44, \citealt{giu+11}; 
HB 21, \citealt{piv+13}; IC 443, \citealt{ack+13}). 
For several low-significance ones (e.g. MSH 17$-$39, \citealt{cas+13}; 
Kes 27, \citealt{xing+15}), \gr\ emission could only be significantly 
detected above $\sim$1 GeV band. The \gr\ spectrum we have obtained
is similar in this respect, although there are only three flux data points and
the uncertainties are large. 

Below we first tested if the \gr\ spectrum could be fitted with a hadronic 
model.  As a leptonic model is often considered for \gr\ emission of SNRs, 
we also tested the leptonic scenario.
We simply assumed that the particles accelerated in the SNR have a power-law 
form with a high-energy cutoff:
\begin{equation}
    dN_i/dE = A_{i}(E/E_0)^{-\alpha_i} \mathrm{exp}(-E/E_{i,c})\ \ \ ,
\end{equation}
where, $i=e,p$, $\alpha_i$ and $E_{i,c}$ are the power-law index and 
high-energy cutoff, respectively. The normalization $A_i$ is determined by 
the total energy ($W_i$) in particles with energies above 1\,GeV.
We used a python package {\it naima} \citep{naima}, a code for computation of 
non-thermal radiation from relativistic particle populations based 
on {\it emcee} \citep{emcee}, to explore the parameter space.

For the hadronic model, the $\gamma$-ray emission is produced by 
the proton-proton inelastic collision.
Due to lack of constraints on the cutoff energy, we fixed $E_{p,c}=1$~PeV 
in our calculation.
Assuming the target density $n_0=1\ {\rm cm^{-3}}$, we obtained 
$\alpha_p=2.83^{+0.43}_{-0.40}$ and $W_p=1.14^{+0.65}_{-0.40}\times10^{51}$ erg.
The resulting best-fit spectral energy distribution (SED) with 
$1\sigma$ confidence interval is shown in the left panel of 
Figure~\ref{fig:sed}.
The fitted proton index is consistent with that of the interacting SNRs 
\citep{acero16}.  Although the energy in protons exceeds the typical explosion 
energy, it is acceptable if the target density is greater than 
$1\, {\rm cm^{-3}}$ (or we may also suspect that the distance of the SNR 
would be significantly smaller than 9.2\,kpc).

\begin{figure}
\centering
\includegraphics[width=0.47\textwidth]{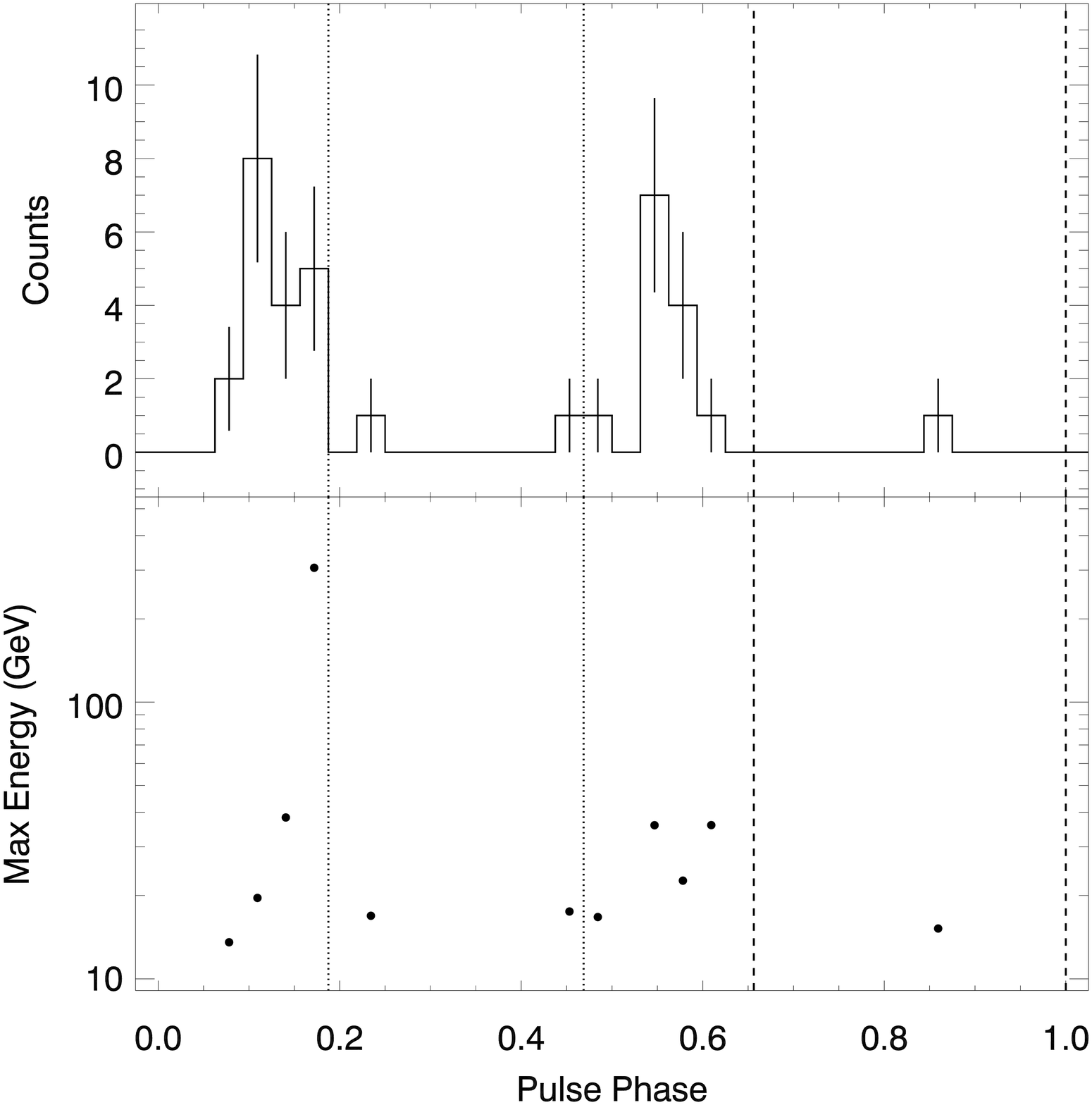}
\caption{Pulse phase distributions of the $>$12 GeV photons ({\it upper} panel) 
and and the maximum photon energies ({\it bottom} panel) 
at the location of PSR~\psr\ (32 phase bins). The dotted and dashed lines 
mark the bridge and offpulse phase intervals, respectively.}
	\label{fig:he}
\end{figure}

The leptonic scenario considers the inverse Compton (IC) process of high-energy
electrons upper-scattering the seed photons in the region.
The cosmic microwave background as well as
the infrared component of the interstellar radiation field (with a temperature
of 35\,K and an energy density of 0.2 ${\rm eV\ cm^{-3}}$;
\citealt{Porter2006,shibata11}) were included in the calculation.
The same population of electrons also produce emission from radio 
to X-rays via the synchrotron process with the average magnetic field $B$.
The resulting best-fit SED with $1\sigma$ confidence interval is displayed 
in the right panel of Figure~\ref{fig:sed}. 
The parameters (with $1\sigma$ uncertainties) are
$\alpha_e=1.82^{+0.11}_{-0.13}$, $E_{e,c}=152^{+202}_{-68}$ GeV, 
$B=2.4^{+2.4}_{-1.0}\ {\rm \mu G}$, 
and $W_e=6.0^{+7.9}_{-3.7}\times10^{49}$ erg.
For the SNR with an age of $\sim10^{5}$~yr \citep{tl06}, the shock velocity 
is expected to decrease below $\sim$100 ${\rm km\,s^{-1}}$. Thus the 
cooling-limited maximum energy of electrons is 
$\sim7(B/2.4\,{\rm \mu G})^{-0.5}$\,TeV \citep[e.g.,][]{zhang2019}.
Considering the higher magnetic field in the early evolution stage, the fitted 
cutoff energy of electrons is in line with expectations.

From the fitting, we found that the SED can be described with reasonable 
parameters of the SNR in either a hadronic or a leptonic scenario. 
The fitting results thus support the association of the \gr\ source PS1 with
the SNR, in addition to their positional coincidence.

\subsection{Origin of the TeV emission?}
\label{sec:tev}

As both HAWC and LHAASO KM2A have detected TeV and upto PeV emissions
respectively in the region of SNR~\snr\ and PSR~\psr\ (while it
can be noted that the two VHE positions have a 
$\sim 2\sigma$ difference) and there are no other potential VHE sources
nearby, it is conceivable to think that the VHE emissions would be
associated with either the SNR or the pulsar. 

Considering that the GeV emission we have detected as arising from the 
SNR~\snr, the extension of the simple power-law GeV spectrum 
would predict lower than the detected TeV--PeV fluxes. 
Also the GeV emission appears away from the HAWC position (with a 2.3$\sigma$
difference). If the VHE emissions are associated with the SNR is thus
highly questionable. While our leptonic modeling does not require
the association as the model fit drops steeply in the VHE range (
cf., Figure~\ref{fig:sed}; we note that unless the one-zone assumption in 
our model is changed), the hadronic model fit is constrained by the
VERITAS upper limits. As shown in the left panel of Figure~\ref{fig:sed}, 
the GeV emission could not be associated with the HAWC source; otherwise
the VHE emission from SNR~\snr\ would have been detected by the VERITAS
observation.  In any case the picture for this region hopefully could be 
clarified with more and deeper VHE observations,
in particular with the LHAASO Water Cherenkov Detector Array (WCDA; 
\citep{lhaaso19})
or the Cherenkov Telescope Array (CTA) north \citep{cta2019.book}. 
The former has a sensitivity 2 times that of HAWC and the latter
would observe the region deeper than VERITAS and potentially detect
the VHE emission from SNR~\snr\ if its emission is due to the hadronic
process.


As PSR~\psr\ is located right within the 1$\sigma$ error circle of 
3HWC~J1954+286, 
their association could be a more likely possibility. 
This pulsar was identified to have pulsed emission detected in $>$10 GeV 
band (\citealt{aje+17}; see also Figure~\ref{fig:sed} and 
Table~\ref{tab:spectra} in this work).
We check the pulse-phase distribution of high energy $\geq 12$\,GeV 
photons in the 0\fdg1 circular region
around the pulsar (note that the LAT's 68\% containment angle at 12\,GeV 
is $\sim 0\fdg1$). As shown in Figure~\ref{fig:he},
their pulse phases match the onpulse phase ranges of the pulsar well.
As a comparison in the bridge and offpulse phase ranges, few such photons
were seen. In addition, the maximum energy of the photons is $\sim 300$\,GeV
(Figure~\ref{fig:he}).
The results do suggest that the pulsar at least have emission upto 
$\sim$100\,GeV. Search for pulsed VHE emission from this pulsar could provide
interesting results, as whether young, energetic pulsars would have VHE
emission similar to the Crab \citep{vcrab11,mcrab12} and the Vela 
pulsar \citep{hvela18} is a question remains to be investigated 
(e.g., \citealt{arc+19} and references therein).

On the other hand as first pointed out by \citet{lin+17}, pulsars
could generally have TeV halos such as the cases of the Geminga and Monogem
pulsars \citep{abe+17}. Using the TeV halo of Geminga as a typical case, 
estimation about the TeV halo of PSR~\psr\  can be made \citep{lin+17}:
its 7\,TeV flux and spatial extension would be
$\sim 1.4\times 10^{-14} (d/1.96\,{\rm kpc})^{-2}$\,TeV$^{-1}$\,cm$^{-2}$\,s$^{-1}$ 
and $\sim 0\fdg2$, respectively, where the distance of 0.19\,kpc and 
the 7\,TeV flux of
$4.9\times 10^{-14}$\,TeV$^{-1}$\,cm$^{-2}$\,s$^{-1}$ for the Geminga are
used. The estimated 7\,TeV flux is approximately 2 times the HAWC measurement
and 3HWC~J1954+286 is given as a point source \citep{3hawc}, but considering
the simple scaling estimation and the uncertainty on
the pulsar's distance, the differences may not be significant.
Given the considerations discussed above, we thus suggest that the VHE 
emissions would more likely be the TeV halo 
associated with the pulsar.

\begin{acknowledgements}
This research made use of the High Performance Computing Resource in the Core
Facility for Advanced Research Computing at Shanghai Astronomical Observatory.
This research was supported by 
the National Natural Science Foundation
of China (11633007, U1738131). Z.W.  acknowledges the support by the Original 
Innovation Program of the Chinese Academy of Sciences (E085021002) and
the Basic Research Program of Yunnan Province No. 202101AS070022. 
X.Z. and Y.C. thank the support of National Key R\&D Program of China under 
Nos. 2018YFA0404204 and 2017YFA0402600 and NSFC grants under Nos. U1931204, 
12173018, and 12121003.
\end{acknowledgements}

\bibliographystyle{aasjournal}

\begin{thebibliography}{}
\expandafter\ifx\csname natexlab\endcsname\relax\def\natexlab#1{#1}\fi
\providecommand{\url}[1]{\href{#1}{#1}}
\providecommand{\dodoi}[1]{doi:~\href{http://doi.org/#1}{\nolinkurl{#1}}}
\providecommand{\doeprint}[1]{\href{http://ascl.net/#1}{\nolinkurl{http://ascl.net/#1}}}
\providecommand{\doarXiv}[1]{\href{https://arxiv.org/abs/#1}{\nolinkurl{https://arxiv.org/abs/#1}}}

\bibitem[{{Abdo} {et~al.}(2009){Abdo}, {Allen}, {Aune}, {Berley}, {Chen},
  {Christopher}, {DeYoung}, {Dingus}, {Ellsworth}, {Gonzalez}, {Goodman},
  {Hays}, {Hoffman}, {H{\"u}ntemeyer}, {Kolterman}, {Linnemann}, {McEnery},
  {Morgan}, {Mincer}, {Nemethy}, {Pretz}, {Ryan}, {Saz Parkinson}, {Shoup},
  {Sinnis}, {Smith}, {Vasileiou}, {Walker}, {Williams}, \&
  {Yodh}}]{abd+09_Milagro}
{Abdo}, A.~A., {Allen}, B.~T., {Aune}, T., {et~al.} 2009, \apjl, 700, L127,
  \dodoi{10.1088/0004-637X/700/2/L127}

\bibitem[{{Abdo} {et~al.}(2013){Abdo}, {Ajello}, {Allafort}, {Baldini},
  {Ballet}, {Barbiellini}, {Baring}, {Bastieri}, {Belfiore}, {Bellazzini}, \&
  et~al.}]{abd+13}
{Abdo}, A.~A., {Ajello}, M., {Allafort}, A., {et~al.} 2013, \apjs, 208, 17,
  \dodoi{10.1088/0067-0049/208/2/17}

\bibitem[{{Abdollahi} {et~al.}(2020){Abdollahi}, {Acero}, {Ackermann},
  {Ajello}, {Atwood}, {Axelsson}, {Baldini}, {Ballet}, {Barbiellini},
  {Bastieri}, {Becerra Gonzalez}, {Bellazzini}, {Berretta}, {Bissaldi}, {Bland
  ford}, {Bloom}, {Bonino}, {Bottacini}, {Brandt}, {Bregeon}, {Bruel},
  {Buehler}, {Burnett}, {Buson}, {Cameron}, {Caputo}, {Caraveo}, {Casandjian},
  {Castro}, {Cavazzuti}, {Charles}, {Chaty}, {Chen}, {Cheung}, {Chiaro},
  {Ciprini}, {Cohen-Tanugi}, {Cominsky}, {Coronado-Bl{\'a}zquez}, {Costantin},
  {Cuoco}, {Cutini}, {D'Ammando}, {DeKlotz}, {Torre Luque}, {de Palma},
  {Desai}, {Digel}, {Lalla}, {Mauro}, {Venere}, {Dom{\'\i}nguez}, {Dumora},
  {Dirirsa}, {Fegan}, {Ferrara}, {Franckowiak}, {Fukazawa}, {Funk}, {Fusco},
  {Gargano}, {Gasparrini}, {Giglietto}, {Giommi}, {Giordano}, {Giroletti},
  {Glanzman}, {Green}, {Grenier}, {Griffin}, {Grondin}, {Grove}, {Guiriec},
  {Harding}, {Hayashi}, {Hays}, {Hewitt}, {Horan}, {J{\'o}hannesson},
  {Johnson}, {Kamae}, {Kerr}, {Kocevski}, {Kovac'evic'}, {Kuss}, {Landriu},
  {Larsson}, {Latronico}, {Lemoine-Goumard}, {Li}, {Liodakis}, {Longo},
  {Loparco}, {Lott}, {Lovellette}, {Lubrano}, {Madejski}, {Maldera},
  {Malyshev}, {Manfreda}, {Marchesini}, {Marcotulli}, {Mart{\'\i}-Devesa},
  {Martin}, {Massaro}, {Mazziotta}, {McEnery}, {Mereu}, {Meyer}, {Michelson},
  {Mirabal}, {Mizuno}, {Monzani}, {Morselli}, {Moskalenko}, {Negro}, {Nuss},
  {Ojha}, {Omodei}, {Orienti}, {Orlando}, {Ormes}, {Palatiello}, {Paliya},
  {Paneque}, {Pei}, {Pe{\~n}a-Herazo}, {Perkins}, {Persic}, {Pesce-Rollins},
  {Petrosian}, {Petrov}, {Piron}, {Poon}, {Porter}, {Principe}, {Rain{\`o}},
  {Rando}, {Razzano}, {Razzaque}, {Reimer}, {Reimer}, {Remy}, {Reposeur},
  {Romani}, {Parkinson}, {Schinzel}, {Serini}, {Sgr{\`o}}, {Siskind}, {Smith},
  {Spandre}, {Spinelli}, {Strong}, {Suson}, {Tajima}, {Takahashi}, {Tak},
  {Thayer}, {Thompson}, {Tibaldo}, {Torres}, {Torresi}, {Valverde}, {Klaveren},
  {Zyl}, {Wood}, {Yassine}, \& {Zaharijas}}]{4fgl20}
{Abdollahi}, S., {Acero}, F., {Ackermann}, M., {et~al.} 2020, \apjs, 247, 33,
  \dodoi{10.3847/1538-4365/ab6bcb}

\bibitem[{{Abeysekara} {et~al.}(2017{\natexlab{a}}){Abeysekara}, {Albert},
  {Alfaro}, {Alvarez}, {{\'A}lvarez}, {Arceo}, {Arteaga-Vel{\'a}zquez}, {Avila
  Rojas}, {Ayala Solares}, {Barber}, {Bautista-Elivar}, {Becerril},
  {Belmont-Moreno}, {BenZvi}, {Berley}, {Bernal}, {Braun}, {Brisbois},
  {Caballero-Mora}, {Capistr{\'a}n}, {Carrami{\~n}ana}, {Casanova}, {Castillo},
  {Cotti}, {Cotzomi}, {Couti{\~n}o de Le{\'o}n}, {De Le{\'o}n}, {De la Fuente},
  {Dingus}, {DuVernois}, {D{\'\i}az-V{\'e}lez}, {Ellsworth}, {Engel},
  {Enr{\'\i}quez-Rivera}, {Fiorino}, {Fraija}, {Garc{\'\i}a-Gonz{\'a}lez},
  {Garfias}, {Gerhardt}, {Gonz{\'a}lez Mu{\~n}oz}, {Gonz{\'a}lez}, {Goodman},
  {Hampel-Arias}, {Harding}, {Hern{\'a}ndez}, {Hern{\'a}ndez-Almada}, {Hinton},
  {Hona}, {Hui}, {H{\"u}ntemeyer}, {Iriarte}, {Jardin-Blicq}, {Joshi},
  {Kaufmann}, {Kieda}, {Lara}, {Lauer}, {Lee}, {Lennarz}, {Vargas},
  {Linnemann}, {Longinotti}, {Luis Raya}, {Luna-Garc{\'\i}a}, {L{\'o}pez-Coto},
  {Malone}, {Marinelli}, {Martinez}, {Martinez-Castellanos},
  {Mart{\'\i}nez-Castro}, {Mart{\'\i}nez-Huerta}, {Matthews},
  {Miranda-Romagnoli}, {Moreno}, {Mostaf{\'a}}, {Nellen}, {Newbold}, {Nisa},
  {Noriega-Papaqui}, {Pelayo}, {Pretz}, {P{\'e}rez-P{\'e}rez}, {Ren}, {Rho},
  {Rivi{\`e}re}, {Rosa-Gonz{\'a}lez}, {Rosenberg}, {Ruiz-Velasco}, {Salazar},
  {Salesa Greus}, {Sandoval}, {Schneider}, {Schoorlemmer}, {Sinnis}, {Smith},
  {Springer}, {Surajbali}, {Taboada}, {Tibolla}, {Tollefson}, {Torres},
  {Ukwatta}, {Vianello}, {Weisgarber}, {Westerhoff}, {Wisher}, {Wood},
  {Yapici}, {Yodh}, {Younk}, {Zepeda}, {Zhou}, {Guo}, {Hahn}, {Li}, \&
  {Zhang}}]{abe+17}
{Abeysekara}, A.~U., {Albert}, A., {Alfaro}, R., {et~al.} 2017{\natexlab{a}},
  Science, 358, 911, \dodoi{10.1126/science.aan4880}

\bibitem[{{Abeysekara} {et~al.}(2017{\natexlab{b}}){Abeysekara}, {Albert},
  {Alfaro}, {Alvarez}, {{\'A}lvarez}, {Arceo}, {Arteaga-Vel{\'a}zquez}, {Ayala
  Solares}, {Barber}, {Baughman}, {Bautista-Elivar}, {Becerra Gonzalez},
  {Becerril}, {Belmont-Moreno}, {BenZvi}, {Berley}, {Bernal}, {Braun},
  {Brisbois}, {Caballero-Mora}, {Capistr{\'a}n}, {Carrami{\~n}ana}, {Casanova},
  {Castillo}, {Cotti}, {Cotzomi}, {Couti{\~n}o de Le{\'o}n}, {de la Fuente},
  {De Le{\'o}n}, {Diaz Hernandez}, {Dingus}, {DuVernois},
  {D{\'\i}az-V{\'e}lez}, {Ellsworth}, {Engel}, {Fiorino}, {Fraija},
  {Garc{\'\i}a-Gonz{\'a}lez}, {Garfias}, {Gerhardt}, {Gonz{\'a}lez Mu{\~n}oz},
  {Gonz{\'a}lez}, {Goodman}, {Hampel-Arias}, {Harding}, {Hernandez},
  {Hernandez-Almada}, {Hinton}, {Hui}, {H{\"u}ntemeyer}, {Iriarte},
  {Jardin-Blicq}, {Joshi}, {Kaufmann}, {Kieda}, {Lara}, {Lauer}, {Lee},
  {Lennarz}, {Le{\'o}n Vargas}, {Linnemann}, {Longinotti}, {Raya},
  {Luna-Garc{\'\i}a}, {L{\'o}pez-Coto}, {Malone}, {Marinelli}, {Martinez},
  {Martinez-Castellanos}, {Mart{\'\i}nez-Castro}, {Mart{\'\i}nez-Huerta},
  {Matthews}, {Miranda-Romagnoli}, {Moreno}, {Mostaf{\'a}}, {Nellen},
  {Newbold}, {Nisa}, {Noriega-Papaqui}, {Pelayo}, {Pretz},
  {P{\'e}rez-P{\'e}rez}, {Ren}, {Rho}, {Rivi{\`e}re}, {Rosa-Gonz{\'a}lez},
  {Rosenberg}, {Ruiz-Velasco}, {Salazar}, {Salesa Greus}, {Sandoval},
  {Schneider}, {Schoorlemmer}, {Sinnis}, {Smith}, {Springer}, {Surajbali},
  {Taboada}, {Tibolla}, {Tollefson}, {Torres}, {Ukwatta}, {Vianello},
  {Villase{\~n}or}, {Weisgarber}, {Westerhoff}, {Wisher}, {Wood}, {Yapici},
  {Younk}, {Zepeda}, \& {Zhou}}]{2hawc}
---. 2017{\natexlab{b}}, \apj, 843, 40, \dodoi{10.3847/1538-4357/aa7556}

\bibitem[{{Acero} {et~al.}(2016){Acero}, {Ackermann}, {Ajello}, {Baldini},
  {Ballet}, {Barbiellini}, {Bastieri}, {Bellazzini}, {Bissaldi}, {Blandford},
  {Bloom}, {Bonino}, {Bottacini}, {Brandt}, {Bregeon}, {Bruel}, {Buehler},
  {Buson}, {Caliandro}, {Cameron}, {Caputo}, {Caragiulo}, {Caraveo},
  {Casandjian}, {Cavazzuti}, {Cecchi}, {Chekhtman}, {Chiang}, {Chiaro},
  {Ciprini}, {Claus}, {Cohen}, {Cohen-Tanugi}, {Cominsky}, {Condon}, {Conrad},
  {Cutini}, {D'Ammando}, {de Angelis}, {de Palma}, {Desiante}, {Digel}, {Di
  Venere}, {Drell}, {Drlica-Wagner}, {Favuzzi}, {Ferrara}, {Franckowiak},
  {Fukazawa}, {Funk}, {Fusco}, {Gargano}, {Gasparrini}, {Giglietto}, {Giommi},
  {Giordano}, {Giroletti}, {Glanzman}, {Godfrey}, {Gomez-Vargas}, {Grenier},
  {Grondin}, {Guillemot}, {Guiriec}, {Gustafsson}, {Hadasch}, {Harding},
  {Hayashida}, {Hays}, {Hewitt}, {Hill}, {Horan}, {Hou}, {Iafrate}, {Jogler},
  {J{\'o}hannesson}, {Johnson}, {Kamae}, {Katagiri}, {Kataoka}, {Katsuta},
  {Kerr}, {Kn{\"o}dlseder}, {Kocevski}, {Kuss}, {Laffon}, {Lande}, {Larsson},
  {Latronico}, {Lemoine-Goumard}, {Li}, {Li}, {Longo}, {Loparco}, {Lovellette},
  {Lubrano}, {Magill}, {Maldera}, {Marelli}, {Mayer}, {Mazziotta}, {Michelson},
  {Mitthumsiri}, {Mizuno}, {Moiseev}, {Monzani}, {Moretti}, {Morselli},
  {Moskalenko}, {Murgia}, {Nemmen}, {Nuss}, {Ohsugi}, {Omodei}, {Orienti},
  {Orlando}, {Ormes}, {Paneque}, {Perkins}, {Pesce-Rollins}, {Petrosian},
  {Piron}, {Pivato}, {Porter}, {Rain{\`o}}, {Rando}, {Razzano}, {Razzaque},
  {Reimer}, {Reimer}, {Renaud}, {Reposeur}, {Rousseau}, {Saz Parkinson},
  {Schmid}, {Schulz}, {Sgr{\`o}}, {Siskind}, {Spada}, {Spandre}, {Spinelli},
  {Strong}, {Suson}, {Tajima}, {Takahashi}, {Tanaka}, {Thayer}, {Thompson},
  {Tibaldo}, {Tibolla}, {Torres}, {Tosti}, {Troja}, {Uchiyama}, {Vianello},
  {Wells}, {Wood}, {Wood}, {Yassine}, {den Hartog}, \& {Zimmer}}]{acero16}
{Acero}, F., {Ackermann}, M., {Ajello}, M., {et~al.} 2016, \apjs, 224, 8,
  \dodoi{10.3847/0067-0049/224/1/8}

\bibitem[{{Ackermann} {et~al.}(2011){Ackermann}, {Ajello}, {Baldini}, {Ballet},
  {Barbiellini}, {Bastieri}, {Bechtol}, {Bellazzini}, {Berenji}, {Bloom},
  {Bonamente}, {Borgland}, {Bouvier}, {Bregeon}, {Brez}, {Brigida}, {Bruel},
  {Buehler}, {Buson}, {Caliandro}, {Cameron}, {Camilo}, {Caraveo},
  {Casandjian}, {Cecchi}, {{\c{C}}elik}, {Charles}, {Chekhtman}, {Cheung},
  {Chiang}, {Ciprini}, {Claus}, {Cognard}, {Cohen-Tanugi}, {Conrad}, {Dermer},
  {de Angelis}, {de Luca}, {de Palma}, {Digel}, {Silva}, {Drell}, {Dubois},
  {Dumora}, {Favuzzi}, {Focke}, {Frailis}, {Fukazawa}, {Funk}, {Fusco},
  {Gargano}, {Germani}, {Giglietto}, {Giommi}, {Giordano}, {Giroletti},
  {Glanzman}, {Godfrey}, {Grenier}, {Grondin}, {Grove}, {Guillemot}, {Guiriec},
  {Hadasch}, {Hanabata}, {Harding}, {Hayashi}, {Hays}, {Hobbs}, {Hughes},
  {J{\'o}hannesson}, {Johnson}, {Johnson}, {Johnston}, {Kamae}, {Katagiri},
  {Kataoka}, {Keith}, {Kerr}, {Kn{\"o}dlseder}, {Kramer}, {Kuss}, {Lande},
  {Latronico}, {Lee}, {Lemoine-Goumard}, {Longo}, {Loparco}, {Lovellette},
  {Lubrano}, {Lyne}, {Makeev}, {Marelli}, {Mazziotta}, {McEnery}, {Mehault},
  {Michelson}, {Mizuno}, {Moiseev}, {Monte}, {Monzani}, {Morselli},
  {Moskalenko}, {Murgia}, {Nakamori}, {Naumann-Godo}, {Nolan}, {Noutsos},
  {Nuss}, {Ohsugi}, {Okumura}, {Ormes}, {Paneque}, {Panetta}, {Parent},
  {Pelassa}, {Pepe}, {Pesce-Rollins}, {Piron}, {Porter}, {Rain{\`o}}, {Rando},
  {Ransom}, {Ray}, {Razzano}, {Rea}, {Reimer}, {Reimer}, {Reposeur}, {Ripken},
  {Ritz}, {Romani}, {Sadrozinski}, {Sander}, {Saz Parkinson}, {Sgr{\`o}},
  {Siskind}, {Smith}, {Smith}, {Spandre}, {Spinelli}, {Strickman}, {Suson},
  {Takahashi}, {Takahashi}, {Tanaka}, {Thayer}, {Thayer}, {Theureau},
  {Thompson}, {Thorsett}, {Tibaldo}, {Torres}, {Tosti}, {Tramacere},
  {Uchiyama}, {Uehara}, {Usher}, {Vandenbroucke}, {Van Etten}, {Vasileiou},
  {Vilchez}, {Vitale}, {Waite}, {Wang}, {Weltevrede}, {Winer}, {Wood}, {Yang},
  {Ylinen}, \& {Ziegler}}]{ack+11}
{Ackermann}, M., {Ajello}, M., {Baldini}, L., {et~al.} 2011, \apj, 726, 35,
  \dodoi{10.1088/0004-637X/726/1/35}

\bibitem[{{Ackermann} {et~al.}(2013){Ackermann}, {Ajello}, {Allafort},
  {Baldini}, {Ballet}, {Barbiellini}, {Baring}, {Bastieri}, {Bechtol},
  {Bellazzini}, {Blandford}, {Bloom}, {Bonamente}, {Borgland}, {Bottacini},
  {Brandt}, {Bregeon}, {Brigida}, {Bruel}, {Buehler}, {Busetto}, {Buson},
  {Caliandro}, {Cameron}, {Caraveo}, {Casandjian}, {Cecchi}, {{\c{C}}elik},
  {Charles}, {Chaty}, {Chaves}, {Chekhtman}, {Cheung}, {Chiang}, {Chiaro},
  {Cillis}, {Ciprini}, {Claus}, {Cohen-Tanugi}, {Cominsky}, {Conrad}, {Corbel},
  {Cutini}, {D'Ammando}, {de Angelis}, {de Palma}, {Dermer}, {do Couto e
  Silva}, {Drell}, {Drlica-Wagner}, {Falletti}, {Favuzzi}, {Ferrara},
  {Franckowiak}, {Fukazawa}, {Funk}, {Fusco}, {Gargano}, {Germani},
  {Giglietto}, {Giommi}, {Giordano}, {Giroletti}, {Glanzman}, {Godfrey},
  {Grenier}, {Grondin}, {Grove}, {Guiriec}, {Hadasch}, {Hanabata}, {Harding},
  {Hayashida}, {Hayashi}, {Hays}, {Hewitt}, {Hill}, {Hughes}, {Jackson},
  {Jogler}, {J{\'o}hannesson}, {Johnson}, {Kamae}, {Kataoka}, {Katsuta},
  {Kn{\"o}dlseder}, {Kuss}, {Lande}, {Larsson}, {Latronico}, {Lemoine-Goumard},
  {Longo}, {Loparco}, {Lovellette}, {Lubrano}, {Madejski}, {Massaro}, {Mayer},
  {Mazziotta}, {McEnery}, {Mehault}, {Michelson}, {Mignani}, {Mitthumsiri},
  {Mizuno}, {Moiseev}, {Monzani}, {Morselli}, {Moskalenko}, {Murgia},
  {Nakamori}, {Nemmen}, {Nuss}, {Ohno}, {Ohsugi}, {Omodei}, {Orienti},
  {Orlando}, {Ormes}, {Paneque}, {Perkins}, {Pesce-Rollins}, {Piron}, {Pivato},
  {Rain{\`o}}, {Rando}, {Razzano}, {Razzaque}, {Reimer}, {Reimer}, {Ritz},
  {Romoli}, {S{\'a}nchez-Conde}, {Schulz}, {Sgr{\`o}}, {Simeon}, {Siskind},
  {Smith}, {Spandre}, {Spinelli}, {Stecker}, {Strong}, {Suson}, {Tajima},
  {Takahashi}, {Takahashi}, {Tanaka}, {Thayer}, {Thayer}, {Thompson},
  {Thorsett}, {Tibaldo}, {Tibolla}, {Tinivella}, {Troja}, {Uchiyama}, {Usher},
  {Vandenbroucke}, {Vasileiou}, {Vianello}, {Vitale}, {Waite}, {Werner},
  {Winer}, {Wood}, {Wood}, {Yamazaki}, {Yang}, \& {Zimmer}}]{ack+13}
{Ackermann}, M., {Ajello}, M., {Allafort}, A., {et~al.} 2013, Science, 339,
  807, \dodoi{10.1126/science.1231160}

\bibitem[{{Ajello} {et~al.}(2017){Ajello}, {Atwood}, {Baldini}, {Ballet},
  {Barbiellini}, {Bastieri}, {Bellazzini}, {Bissaldi}, {Blandford}, {Bloom},
  {Bonino}, {Bregeon}, {Britto}, {Bruel}, {Buehler}, {Buson}, {Cameron},
  {Caputo}, {Caragiulo}, {Caraveo}, {Cavazzuti}, {Cecchi}, {Charles},
  {Chekhtman}, {Cheung}, {Chiaro}, {Ciprini}, {Cohen}, {Costantin}, {Costanza},
  {Cuoco}, {Cutini}, {D'Ammando}, {de Palma}, {Desiante}, {Digel}, {Di Lalla},
  {Di Mauro}, {Di Venere}, {Dom{\'\i}nguez}, {Drell}, {Dumora}, {Favuzzi},
  {Fegan}, {Ferrara}, {Fortin}, {Franckowiak}, {Fukazawa}, {Funk}, {Fusco},
  {Gargano}, {Gasparrini}, {Giglietto}, {Giommi}, {Giordano}, {Giroletti},
  {Glanzman}, {Green}, {Grenier}, {Grondin}, {Grove}, {Guillemot}, {Guiriec},
  {Harding}, {Hays}, {Hewitt}, {Horan}, {J{\'o}hannesson}, {Kensei}, {Kuss},
  {La Mura}, {Larsson}, {Latronico}, {Lemoine-Goumard}, {Li}, {Longo},
  {Loparco}, {Lott}, {Lubrano}, {Magill}, {Maldera}, {Manfreda}, {Mazziotta},
  {McEnery}, {Meyer}, {Michelson}, {Mirabal}, {Mitthumsiri}, {Mizuno},
  {Moiseev}, {Monzani}, {Morselli}, {Moskalenko}, {Negro}, {Nuss}, {Ohsugi},
  {Omodei}, {Orienti}, {Orlando}, {Palatiello}, {Paliya}, {Paneque}, {Perkins},
  {Persic}, {Pesce-Rollins}, {Piron}, {Porter}, {Principe}, {Rain{\`o}},
  {Rando}, {Razzano}, {Razzaque}, {Reimer}, {Reimer}, {Reposeur}, {Saz
  Parkinson}, {Sgr{\`o}}, {Simone}, {Siskind}, {Spada}, {Spandre}, {Spinelli},
  {Stawarz}, {Suson}, {Takahashi}, {Tak}, {Thayer}, {Thayer}, {Thompson},
  {Torres}, {Torresi}, {Troja}, {Vianello}, {Wood}, \& {Wood}}]{aje+17}
{Ajello}, M., {Atwood}, W.~B., {Baldini}, L., {et~al.} 2017, \apjs, 232, 18,
  \dodoi{10.3847/1538-4365/aa8221}

\bibitem[{{Albert} {et~al.}(2020){Albert}, {Alfaro}, {Alvarez}, {Camacho},
  {Arteaga-Vel{\'a}zquez}, {Arunbabu}, {Avila Rojas}, {Ayala Solares},
  {Baghmanyan}, {Belmont-Moreno}, {BenZvi}, {Brisbois}, {Caballero-Mora},
  {Capistr{\'a}n}, {Carrami{\~n}ana}, {Casanova}, {Cotti}, {Couti{\~n}o de
  Le{\'o}n}, {De la Fuente}, {Diaz Hernandez}, {Diaz-Cruz}, {Dingus},
  {DuVernois}, {Durocher}, {D{\'\i}az-V{\'e}lez}, {Ellsworth}, {Engel},
  {Espinoza}, {Fan}, {Fang}, {Alonso}, {Fleischhack}, {Fraija},
  {Galv{\'a}n-G{\'a}mez}, {Garcia}, {Garc{\'\i}a-Gonz{\'a}lez}, {Garfias},
  {Giacinti}, {Gonz{\'a}lez}, {Goodman}, {Harding}, {Hernandez}, {Hinton},
  {Hona}, {Huang}, {Hueyotl-Zahuantitla}, {H{\"u}ntemeyer}, {Iriarte},
  {Jardin-Blicq}, {Joshi}, {Kieda}, {Lara}, {Lee}, {Le{\'o}n Vargas},
  {Linnemann}, {Longinotti}, {Luis-Raya}, {Lundeen}, {L{\'o}pez-Coto},
  {Malone}, {Marandon}, {Martinez}, {Martinez-Castellanos},
  {Mart{\'\i}nez-Castro}, {Matthews}, {Miranda-Romagnoli}, {Morales-Soto},
  {Moreno}, {Mostaf{\'a}}, {Nayerhoda}, {Nellen}, {Newbold}, {Nisa},
  {Noriega-Papaqui}, {Olivera-Nieto}, {Omodei}, {Peisker}, {P{\'e}rez Araujo},
  {P{\'e}rez-P{\'e}rez}, {Ren}, {Rho}, {Rivi{\`e}re}, {Rosa-Gonz{\'a}lez},
  {Ruiz-Velasco}, {Salazar}, {Salesa Greus}, {Sandoval}, {Schneider},
  {Schoorlemmer}, {Serna}, {Sinnis}, {Smith}, {Springer}, {Surajbali},
  {Tollefson}, {Torres}, {Torres-Escobedo}, {Ukwatta}, {Ure{\~n}a-Mena},
  {Weisgarber}, {Werner}, {Willox}, {Zepeda}, {Zhou}, {de Le{\'o}n},
  {{\'A}lvarez}, \& {HAWC Collaboration}}]{3hawc}
{Albert}, A., {Alfaro}, R., {Alvarez}, C., {et~al.} 2020, \apj, 905, 76,
  \dodoi{10.3847/1538-4357/abc2d8}

\bibitem[{{Aleksi{\'c}} {et~al.}(2012){Aleksi{\'c}}, {Alvarez}, {Antonelli},
  {Antoranz}, {Asensio}, {Backes}, {Barrio}, {Bastieri}, {Becerra
  Gonz{\'a}lez}, {Bednarek}, {Berdyugin}, {Berger}, {Bernardini}, {Biland},
  {Blanch}, {Bock}, {Boller}, {Bonnoli}, {Borla Tridon}, {Braun}, {Bretz},
  {Ca{\~n}ellas}, {Carmona}, {Carosi}, {Colin}, {Colombo}, {Contreras},
  {Cortina}, {Cossio}, {Covino}, {Dazzi}, {de Angelis}, {de Caneva}, {de Cea
  Del Pozo}, {de Lotto}, {Delgado Mendez}, {Diago Ortega}, {Doert},
  {Dom{\'\i}nguez}, {Dominis Prester}, {Dorner}, {Doro}, {Eisenacher},
  {Elsaesser}, {Ferenc}, {Fonseca}, {Font}, {Fruck}, {Garc{\'\i}a L{\'o}pez},
  {Garczarczyk}, {Garrido}, {Giavitto}, {Godinovi{\'c}}, {Hadasch},
  {H{\"a}fner}, {Herrero}, {Hildebrand}, {H{\"o}hne-M{\"o}nch}, {Hose},
  {Hrupec}, {Jogler}, {Kellermann}, {Klepser}, {Kr{\"a}henb{\"u}hl}, {Krause},
  {Kushida}, {La Barbera}, {Lelas}, {Leonardo}, {Lewandowska}, {Lindfors},
  {Lombardi}, {L{\'o}pez}, {L{\'o}pez-Oramas}, {Lorenz}, {Makariev}, {Maneva},
  {Mankuzhiyil}, {Mannheim}, {Maraschi}, {Mariotti}, {Mart{\'\i}nez}, {Mazin},
  {Meucci}, {Miranda}, {Mirzoyan}, {Mold{\'o}n}, {Moralejo}, {Munar-Adrover},
  {Niedzwiecki}, {Nieto}, {Nilsson}, {Nowak}, {Orito}, {Paneque}, {Paoletti},
  {Pardo}, {Paredes}, {Partini}, {Perez-Torres}, {Persic}, {Peruzzo}, {Pilia},
  {Pochon}, {Prada}, {Prada Moroni}, {Prandini}, {Puerto Gimenez}, {Puljak},
  {Reichardt}, {Reinthal}, {Rhode}, {Rib{\'o}}, {Rico}, {R{\"u}gamer},
  {Saggion}, {Saito}, {Saito}, {Salvati}, {Satalecka}, {Scalzotto}, {Scapin},
  {Schultz}, {Schweizer}, {Shayduk}, {Shore}, {Sillanp{\"a}{\"a}}, {Sitarek},
  {{\v{S}}nidari{\'c}}, {Sobczynska}, {Spanier}, {Spiro}, {Stamatescu},
  {Stamerra}, {Steinke}, {Storz}, {Strah}, {Suri{\'c}}, {Takalo}, {Takami},
  {Tavecchio}, {Temnikov}, {Terzi{\'c}}, {Tescaro}, {Teshima}, {Tibolla},
  {Torres}, {Treves}, {Uellenbeck}, {Vankov}, {Vogler}, {Wagner}, {Weitzel},
  {Zabalza}, {Zandanel}, {Zanin}, \& {Hirotani}}]{mcrab12}
{Aleksi{\'c}}, J., {Alvarez}, E.~A., {Antonelli}, L.~A., {et~al.} 2012, \aap,
  540, A69, \dodoi{10.1051/0004-6361/201118166}

\bibitem[{{Archer} {et~al.}(2019){Archer}, {Benbow}, {Bird}, {Brose},
  {Buchovecky}, {Buckley}, {Chromey}, {Cui}, {Falcone}, {Feng}, {Finley},
  {Fortson}, {Furniss}, {Gent}, {Gueta}, {Hanna}, {Hassan}, {Hervet}, {Holder},
  {Hughes}, {Humensky}, {Johnson}, {Kaaret}, {Kar}, {Kelley-Hoskins},
  {Kertzman}, {Kieda}, {Krennrich}, {Kumar}, {Lang}, {Lin}, {McCann},
  {Moriarty}, {Mukherjee}, {O'Brien}, {Ong}, {Otte}, {Pandel}, {Park},
  {Petrashyk}, {Pohl}, {Pueschel}, {Quinn}, {Ragan}, {Richards}, {Roache},
  {Sadeh}, {Santander}, {Scott}, {Sembroski}, {Shahinyan}, {Sushch}, {Tyler},
  {Wakely}, {Weinstein}, {Wells}, {Wilcox}, {Wilhelm}, {Williams},
  {Williamson}, \& {Zitzer}}]{arc+19}
{Archer}, A., {Benbow}, W., {Bird}, R., {et~al.} 2019, \apj, 876, 95,
  \dodoi{10.3847/1538-4357/ab14f4}

\bibitem[{{Atwood} {et~al.}(2009){Atwood}, {Abdo}, {Ackermann}, {Althouse},
  {Anderson}, {Axelsson}, {Baldini}, {Ballet}, {Band}, {Barbiellini}, \&
  et~al.}]{atw+09}
{Atwood}, W.~B., {Abdo}, A.~A., {Ackermann}, M., {et~al.} 2009, \apj, 697,
  1071, \dodoi{10.1088/0004-637X/697/2/1071}

\bibitem[{{Bai} {et~al.}(2019){Bai}, {Bi}, {Bi}, {Cao}, {Chen}, {Chen},
  {Chiavassa}, {Cui}, {Dai}, {della Volpe}, {Di Girolamo}, {Di Sciascio},
  {Fan}, {Giacalone}, {Guo}, {He}, {He}, {Heller}, {Huang}, {Huang}, {Jia},
  {Ksenofontov}, {Leahy}, {Li}, {Li}, {Liang}, {Lipari}, {Liu}, {Liu}, {Liu},
  {Ma}, {Martineau-Huynh}, {Martraire}, {Montaruli}, {Ruffolo}, {Stenkin},
  {Su}, {Tam}, {Tang}, {Tian}, {Vallania}, {Vernetto}, {Vigorito}, {Wang},
  {Wang}, {Wang}, {Wang}, {Wang}, {Wang}, {Wei}, {Wei}, {Wu}, {Wu}, {Wu},
  {Yan}, {Yang}, {Yang}, {Yao}, {Yin}, {Yuan}, {Zhang}, {Zhang}, {Zhang},
  {Zhang}, {Zhang}, {Zhang}, {Zhao}, {Zhou}, {Zhu}, \& {Zhu}}]{lhaaso19}
{Bai}, X., {Bi}, B.~Y., {Bi}, X.~J., {et~al.} 2019, arXiv e-prints,
  arXiv:1905.02773.
\newblock \doarXiv{1905.02773}

\bibitem[{{Ballet} {et~al.}(2020){Ballet}, {Burnett}, {Digel}, \&
  {Lott}}]{bal+20}
{Ballet}, J., {Burnett}, T.~H., {Digel}, S.~W., \& {Lott}, B. 2020, arXiv
  e-prints, arXiv:2005.11208.
\newblock \doarXiv{2005.11208}

\bibitem[{{Cao} {et~al.}(2021){Cao}, {Aharonian}, {An}, {Axikegu}, {Bai},
  {Bao}, {Bastieri}, {Bi}, {Bi}, {Cai}, {Cai}, {Cao}, {Chang}, {Chang},
  {Chang}, {Chen}, {Chen}, {Chen}, {Chen}, {Chen}, {Chen}, {Chen}, {Chen},
  {Chen}, {Chen}, {Chen}, {Chen}, {Chen}, {Cheng}, {Cheng}, {Cui}, {Cui},
  {Cui}, {Dai}, {Dai}, {Dai}, {Danzengluobu}, {della Volpe}, {D'Ettorre
  Piazzoli}, {Dong}, {Fan}, {Fan}, {Fan}, {Fang}, {Fang}, {Feng}, {Feng},
  {Feng}, {Feng}, {Gao}, {Gao}, {Gao}, {Gao}, {Ge}, {Geng}, {Gong}, {Gou},
  {Gu}, {Guo}, {Guo}, {Guo}, {Guo}, {Han}, {He}, {He}, {He}, {He}, {He}, {He},
  {Heller}, {Hor}, {Hou}, {Hou}, {Hu}, {Hu}, {Hu}, {Hu}, {Huang}, {Huang},
  {Huang}, {Huang}, {Huang}, {Ji}, {Ji}, {Jia}, {Jiang}, {Jiang}, {Jin},
  {Kuleshov}, {Levochkin}, {Li}, {Li}, {Li}, {Li}, {Li}, {Li}, {Li}, {Li},
  {Li}, {Li}, {Li}, {Li}, {Li}, {Li}, {Li}, {Li}, {Li}, {Liang}, {Liang},
  {Lin}, {Liu}, {Liu}, {Liu}, {Liu}, {Liu}, {Liu}, {Liu}, {Liu}, {Liu}, {Liu},
  {Liu}, {Liu}, {Liu}, {Liu}, {Liu}, {Long}, {Lu}, {Lv}, {Ma}, {Ma}, {Ma},
  {Mao}, {Masood}, {Mitthumsiri}, {Montaruli}, {Nan}, {Pang},
  {Pattarakijwanich}, {Pei}, {Qi}, {Ruffolo}, {Rulev}, {S{\'a}iz}, {Shao},
  {Shchegolev}, {Sheng}, {Shi}, {Song}, {Stenkin}, {Stepanov}, {Sun}, {Sun},
  {Sun}, {Tam}, {Tang}, {Tian}, {Wang}, {Wang}, {Wang}, {Wang}, {Wang}, {Wang},
  {Wang}, {Wang}, {Wang}, {Wang}, {Wang}, {Wang}, {Wang}, {Wang}, {Wang},
  {Wang}, {Wang}, {Wang}, {Wang}, {Wang}, {Wang}, {Wei}, {Wei}, {Wei}, {Wen},
  {Wu}, {Wu}, {Wu}, {Wu}, {Wu}, {Xi}, {Xia}, {Xia}, {Xiang}, {Xiao}, {Xiao},
  {Xin}, {Xin}, {Xing}, {Xu}, {Xu}, {Xue}, {Yan}, {Yang}, {Yang}, {Yang},
  {Yang}, {Yang}, {Yang}, {Yang}, {Yao}, {Yao}, {Ye}, {Yin}, {Yin}, {You},
  {You}, {Yu}, {Yuan}, {Zeng}, {Zeng}, {Zeng}, {Zeng}, {Zha}, {Zhai}, {Zhang},
  {Zhang}, {Zhang}, {Zhang}, {Zhang}, {Zhang}, {Zhang}, {Zhang}, {Zhang},
  {Zhang}, {Zhang}, {Zhang}, {Zhang}, {Zhang}, {Zhang}, {Zhang}, {Zhang},
  {Zhang}, {Zhang}, {Zhao}, {Zhao}, {Zhao}, {Zhao}, {Zhao}, {Zheng}, {Zheng},
  {Zhou}, {Zhou}, {Zhou}, {Zhou}, {Zhou}, {Zhou}, {Zhu}, {Zhu}, {Zhu}, {Zhu},
  \& {Zuo}}]{lhaaso_pev_2021}
{Cao}, Z., {Aharonian}, F.~A., {An}, Q., {et~al.} 2021, \nat, 594, 33,
  \dodoi{10.1038/s41586-021-03498-z}

\bibitem[{{Castro} {et~al.}(2013){Castro}, {Slane}, {Carlton}, \&
  {Figueroa-Feliciano}}]{cas+13}
{Castro}, D., {Slane}, P., {Carlton}, A., \& {Figueroa-Feliciano}, E. 2013,
  \apj, 774, 36, \dodoi{10.1088/0004-637X/774/1/36}

\bibitem[{{Cherenkov Telescope Array Consortium} {et~al.}(2019){Cherenkov
  Telescope Array Consortium}, {Acharya}, {Agudo}, {Al Samarai}, {Alfaro},
  {Alfaro}, {Alispach}, {Alves Batista}, {Amans}, {Amato}, {Ambrosi},
  {Antolini}, {Antonelli}, {Aramo}, {Araya}, {Armstrong}, {Arqueros},
  {Arrabito}, {Asano}, {Ashley}, {Backes}, {Balazs}, {Balbo}, {Ballester},
  {Ballet}, {Bamba}, {Barkov}, {Barres de Almeida}, {Barrio}, {Bastieri},
  {Becherini}, {Belfiore}, {Benbow}, {Berge}, {Bernardini}, {Bernardini},
  {Bernardos}, {Bernl{\"o}hr}, {Bertucci}, {Biasuzzi}, {Bigongiari}, {Biland},
  {Bissaldi}, {Biteau}, {Blanch}, {Blazek}, {Boisson}, {Bolmont}, {Bonanno},
  {Bonardi}, {Bonavolont{\`a}}, {Bonnoli}, {Bosnjak}, {B{\"o}ttcher},
  {Braiding}, {Bregeon}, {Brill}, {Brown}, {Brun}, {Brunetti}, {Buanes},
  {Buckley}, {Bugaev}, {B{\"u}hler}, {Bulgarelli}, {Bulik}, {Burton},
  {Burtovoi}, {Busetto}, {Canestrari}, {Capalbi}, {Capitanio}, {Caproni},
  {Caraveo}, {C{\'a}rdenas}, {Carlile}, {Carosi}, {Carqu{\'\i}n}, {Carr},
  {Casanova}, {Cascone}, {Catalani}, {Catalano}, {Cauz}, {Cerruti}, {Chadwick},
  {Chaty}, {Chaves}, {Chen}, {Chen}, {Chernyakova}, {Chikawa}, {Christov},
  {Chudoba}, {Cie{\'s}lar}, {Coco}, {Colafrancesco}, {Colin}, {Conforti},
  {Connaughton}, {Conrad}, {Contreras}, {Cortina}, {Costa}, {Costantini},
  {Cotter}, {Covino}, {Crocker}, {Cuadra}, {Cuevas}, {Cumani}, {D'A{\`\i}},
  {D'Ammando}, {D'Avanzo}, {D'Urso}, {Daniel}, {Davids}, {Dawson}, {Dazzi}, {De
  Angelis}, {de C{\'a}ssia dos Anjos}, {De Cesare}, {De Franco}, {de Gouveia
  Dal Pino}, {de la Calle}, {de los Reyes Lopez}, {De Lotto}, {De Luca}, {De
  Lucia}, {de Naurois}, {de O{\~n}a Wilhelmi}, {De Palma}, {De Persio}, {de
  Souza}, {Deil}, {Del Santo}, {Delgado}, {della Volpe}, {Di Girolamo}, {Di
  Pierro}, {Di Venere}, {D{\'\i}az}, {Dib}, {Diebold}, {Djannati-Ata{\"\i}},
  {Dom{\'\i}nguez}, {Dominis Prester}, {Dorner}, {Doro}, {Drass}, {Dravins},
  {Dubus}, {Dwarkadas}, {Ebr}, {Eckner}, {Egberts}, {Einecke}, {Ekoume},
  {Els{\"a}sser}, {Ernenwein}, {Espinoza}, {Evoli}, {Fairbairn},
  {Falceta-Goncalves}, {Falcone}, {Farnier}, {Fasola}, {Fedorova}, {Fegan},
  {Fernandez-Alonso}, {Fern{\'a}ndez-Barral}, {Ferrand}, {Fesquet},
  {Filipovic}, {Fioretti}, {Fontaine}, {Fornasa}, {Fortson}, {Freixas
  Coromina}, {Fruck}, {Fujita}, {Fukazawa}, {Funk}, {F{\"u}{\ss}ling},
  {Gabici}, {Gadola}, {Gallant}, {Garcia}, {Garcia L{\'o}pez}, {Garczarczyk},
  {Gaskins}, {Gasparetto}, {Gaug}, {Gerard}, {Giavitto}, {Giglietto}, {Giommi},
  {Giordano}, {Giro}, {Giroletti}, {Giuliani}, {Glicenstein}, {Gnatyk},
  {Godinovic}, {Goldoni}, {G{\'o}mez-Vargas}, {Gonz{\'a}lez}, {Gonz{\'a}lez},
  {G{\"o}tz}, {Graham}, {Grandi}, {Granot}, {Green}, {Greenshaw}, {Griffiths},
  {Gunji}, {Hadasch}, {Hara}, {Hardcastle}, {Hassan}, {Hayashi}, {Hayashida},
  {Heller}, {Helo}, {Hermann}, {Hinton}, {Hnatyk}, {Hofmann}, {Holder},
  {Horan}, {H{\"o}randel}, {Horns}, {Horvath}, {Hovatta}, {Hrabovsky},
  {Hrupec}, {Humensky}, {H{\"u}tten}, {Iarlori}, {Inada}, {Inome}, {Inoue},
  {Inoue}, {Inoue}, {Iocco}, {Ioka}, {Iori}, {Ishio}, {Iwamura}, {Jamrozy},
  {Janecek}, {Jankowsky}, {Jean}, {Jung-Richardt}, {Jurysek}, {Kaaret},
  {Karkar}, {Katagiri}, {Katz}, {Kawanaka}, {Kazanas}, {Kh{\'e}lifi}, {Kieda},
  {Kimeswenger}, {Kimura}, {Kisaka}, {Knapp}, {Kn{\"o}dlseder}, {Koch},
  {Kohri}, {Komin}, {Kosack}, {Kraus}, {Krause}, {Krau{\ss}}, {Kubo}, {Kukec
  Mezek}, {Kuroda}, {Kushida}, {La Palombara}, {Lamanna}, {Lang}, {Lapington},
  {Le Blanc}, {Leach}, {Lees}, {Lefaucheur}, {Leigui de Oliveira}, {Lenain},
  {Lico}, {Limon}, {Lindfors}, {Lohse}, {Lombardi}, {Longo}, {L{\'o}pez},
  {L{\'o}pez-Coto}, {Lu}, {Lucarelli}, {Luque-Escamilla}, {Lyard}, {Maccarone},
  {Maier}, {Majumdar}, {Malaguti}, {Mandat}, {Maneva}, {Manganaro}, {Mangano},
  {Marcowith}, {Mar{\'\i}n}, {Markoff}, {Mart{\'\i}}, {Martin},
  {Mart{\'\i}nez}, {Mart{\'\i}nez}, {Masetti}, {Masuda}, {Maurin}, {Maxted},
  {Mazin}, {Medina}, {Melandri}, {Mereghetti}, {Meyer}, {Minaya}, {Mirabal},
  {Mirzoyan}, {Mitchell}, {Mizuno}, {Moderski}, {Mohammed}, {Mohrmann},
  {Montaruli}, {Moralejo}, {Morcuende-Parrilla}, {Mori}, {Morlino}, {Morris},
  {Morselli}, {Moulin}, {Mukherjee}, {Mundell}, {Murach}, {Muraishi}, {Murase},
  {Nagai}, {Nagataki}, {Nagayoshi}, {Naito}, {Nakamori}, {Nakamura}, {Niemiec},
  {Nieto}, {Niko{\l}ajuk}, {Nishijima}, {Noda}, {Nosek}, {Novosyadlyj},
  {Nozaki}, {O'Brien}, {Oakes}, {Ohira}, {Ohishi}, {Ohm}, {Okazaki}, {Okumura},
  {Ong}, {Orienti}, {Orito}, {Osborne}, {Ostrowski}, {Otte}, {Oya}, {Padovani},
  {Paizis}, {Palatiello}, {Palatka}, {Paoletti}, {Paredes}, {Pareschi},
  {Parsons}, {Pe'er}, {Pech}, {Pedaletti}, {Perri}, {Persic}, {Petrashyk},
  {Petrucci}, {Petruk}, {Peyaud}, {Pfeifer}, {Piano}, {Pisarski}, {Pita},
  {Pohl}, {Polo}, {Pozo}, {Prandini}, {Prast}, {Principe}, {Prokhorov},
  {Prokoph}, {Prouza}, {P{\"u}hlhofer}, {Punch}, {P{\"u}rckhauer}, {Queiroz},
  {Quirrenbach}, {Rain{\`o}}, {Razzaque}, {Reimer}, {Reimer}, {Reisenegger},
  {Renaud}, {Rezaeian}, {Rhode}, {Ribeiro}, {Rib{\'o}}, {Richtler}, {Rico},
  {Rieger}, {Riquelme}, {Rivoire}, {Rizi}, {Rodriguez}, {Rodriguez Fernandez},
  {Rodr{\'\i}guez V{\'a}zquez}, {Rojas}, {Romano}, {Romeo}, {Rosado}, {Rovero},
  {Rowell}, {Rudak}, {Rugliancich}, {Rulten}, {Sadeh}, {Safi-Harb}, {Saito},
  {Sakaki}, {Sakurai}, {Salina}, {S{\'a}nchez-Conde}, {Sandaker}, {Sandoval},
  {Sangiorgi}, {Sanguillon}, {Sano}, {Santander}, {Sarkar}, {Satalecka},
  {Saturni}, {Schioppa}, {Schlenstedt}, {Schneider}, {Schoorlemmer},
  {Schovanek}, {Schulz}, {Schussler}, {Schwanke}, {Sciacca}, {Scuderi},
  {Seitenzahl}, {Semikoz}, {Sergijenko}, {Servillat}, {Shalchi}, {Shellard},
  {Sidoli}, {Siejkowski}, {Sillanp{\"a}{\"a}}, {Sironi}, {Sitarek}, {Sliusar},
  {Slowikowska}, {Sol}, {Stamerra}, {Stani{\v{c}}}, {Starling}, {Stawarz},
  {Stefanik}, {Stephan}, {Stolarczyk}, {Stratta}, {Straumann}, {Suomijarvi},
  {Supanitsky}, {Tagliaferri}, {Tajima}, {Tavani}, {Tavecchio}, {Tavernet},
  {Tayabaly}, {Tejedor}, {Temnikov}, {Terada}, {Terrier}, {Terzic}, {Teshima},
  {Testa}, {Thoudam}, {Tian}, {Tibaldo}, {Tluczykont}, {Todero Peixoto},
  {Tokanai}, {Tomastik}, {Tonev}, {Tornikoski}, {Torres}, {Torresi}, {Tosti},
  {Tothill}, {Tovmassian}, {Travnicek}, {Trichard}, {Trifoglio}, {Troyano
  Pujadas}, {Tsujimoto}, {Umana}, {Vagelli}, {Vagnetti}, {Valentino},
  {Vallania}, {Valore}, {van Eldik}, {Vandenbroucke}, {Varner}, {Vasileiadis},
  {Vassiliev}, {V{\'a}zquez Acosta}, {Vecchi}, {Vega}, {Vercellone}, {Veres},
  {Vergani}, {Verzi}, {Vettolani}, {Viana}, {Vigorito}, {Villanueva}, {Voelk},
  {Vollhardt}, {Vorobiov}, {Vrastil}, {Vuillaume}, {Wagner}, {Wagner},
  {Walter}, {Ward}, {Warren}, {Watson}, {Werner}, {White}, {White},
  {Wierzcholska}, {Wilcox}, {Will}, {Williams}, {Wischnewski}, {Wood},
  {Yamamoto}, {Yamazaki}, {Yanagita}, {Yang}, {Yoshida}, {Yoshiike},
  {Yoshikoshi}, {Zacharias}, {Zaharijas}, {Zampieri}, {Zandanel}, {Zanin},
  {Zavrtanik}, {Zavrtanik}, {Zdziarski}, {Zech}, {Zechlin}, {Zhdanov},
  {Ziegler}, \& {Zorn}}]{cta2019.book}
{Cherenkov Telescope Array Consortium}, {Acharya}, B.~S., {Agudo}, I., {et~al.}
  2019, {Science with the Cherenkov Telescope Array} (World Scientific
  Publishing Co. Pte. Ltd.), \dodoi{10.1142/10986}

\bibitem[{{Edwards} {et~al.}(2006){Edwards}, {Hobbs}, \&
  {Manchester}}]{edw2006}
{Edwards}, R.~T., {Hobbs}, G.~B., \& {Manchester}, R.~N. 2006, \mnras, 372,
  1549, \dodoi{10.1111/j.1365-2966.2006.10870.x}

\bibitem[{{Foreman-Mackey} {et~al.}(2013){Foreman-Mackey}, {Hogg}, {Lang}, \&
  {Goodman}}]{emcee}
{Foreman-Mackey}, D., {Hogg}, D.~W., {Lang}, D., \& {Goodman}, J. 2013, \pasp,
  125, 306, \dodoi{10.1086/670067}

\bibitem[{{Gao} {et~al.}(2011){Gao}, {Han}, {Reich}, {Reich}, {Sun}, \&
  {Xiao}}]{gao+11}
{Gao}, X.~Y., {Han}, J.~L., {Reich}, W., {et~al.} 2011, \aap, 529, A159,
  \dodoi{10.1051/0004-6361/201016311}

\bibitem[{{Giuliani} {et~al.}(2011){Giuliani}, {Cardillo}, {Tavani}, {Fukui},
  {Yoshiike}, {Torii}, {Dubner}, {Castelletti}, {Barbiellini}, {Bulgarelli},
  {Caraveo}, {Costa}, {Cattaneo}, {Chen}, {Contessi}, {Del Monte},
  {Donnarumma}, {Evangelista}, {Feroci}, {Gianotti}, {Lazzarotto}, {Lucarelli},
  {Longo}, {Marisaldi}, {Mereghetti}, {Pacciani}, {Pellizzoni}, {Piano},
  {Picozza}, {Pittori}, {Pucella}, {Rapisarda}, {Rappoldi}, {Sabatini},
  {Soffitta}, {Striani}, {Trifoglio}, {Trois}, {Vercellone}, {Verrecchia},
  {Vittorini}, {Colafrancesco}, {Giommi}, \& {Bignami}}]{giu+11}
{Giuliani}, A., {Cardillo}, M., {Tavani}, M., {et~al.} 2011, \apjl, 742, L30,
  \dodoi{10.1088/2041-8205/742/2/L30}

\bibitem[{{H.~E.~S.~S. Collaboration} {et~al.}(2018{\natexlab{a}}){H.~E.~S.~S.
  Collaboration}, {Abdalla}, {Abramowski}, {Aharonian}, {Ait Benkhali},
  {Ang{\"u}ner}, {Arakawa}, {Arrieta}, {Aubert}, {Backes}, {Balzer}, {Barnard},
  {Becherini}, {Becker Tjus}, {Berge}, {Bernhard}, {Bernl{\"o}hr}, {Blackwell},
  {B{\"o}ttcher}, {Boisson}, {Bolmont}, {Bonnefoy}, {Bordas}, {Bregeon},
  {Brun}, {Brun}, {Bryan}, {B{\"u}chele}, {Bulik}, {Capasso}, {Carrigan},
  {Caroff}, {Carosi}, {Casanova}, {Cerruti}, {Chakraborty}, {Chaves}, {Chen},
  {Chevalier}, {Colafrancesco}, {Condon}, {Conrad}, {Davids}, {Decock}, {Deil},
  {Devin}, {deWilt}, {Dirson}, {Djannati-Ata{\"\i}}, {Domainko}, {Donath},
  {Drury}, {Dutson}, {Dyks}, {Edwards}, {Egberts}, {Eger}, {Emery},
  {Ernenwein}, {Eschbach}, {Farnier}, {Fegan}, {Fernandes}, {Fiasson},
  {Fontaine}, {F{\"o}rster}, {Funk}, {F{\"u}{\ss}ling}, {Gabici}, {Gallant},
  {Garrigoux}, {Gast}, {Gat{\'e}}, {Giavitto}, {Giebels}, {Glawion},
  {Glicenstein}, {Gottschall}, {Grondin}, {Hahn}, {Haupt}, {Hawkes},
  {Heinzelmann}, {Henri}, {Hermann}, {Hinton}, {Hofmann}, {Hoischen}, {Holch},
  {Holler}, {Horns}, {Ivascenko}, {Iwasaki}, {Jacholkowska}, {Jamrozy},
  {Jankowsky}, {Jankowsky}, {Jingo}, {Jouvin}, {Jung-Richardt}, {Kastendieck},
  {Katarzy{\'n}ski}, {Katsuragawa}, {Katz}, {Kerszberg}, {Khangulyan},
  {Kh{\'e}lifi}, {King}, {Klepser}, {Klochkov}, {Klu{\'z}niak}, {Komin},
  {Kosack}, {Krakau}, {Kraus}, {Kr{\"u}ger}, {Laffon}, {Lamanna}, {Lau},
  {Lees}, {Lefaucheur}, {Lemi{\`e}re}, {Lemoine-Goumard}, {Lenain}, {Leser},
  {Lohse}, {Lorentz}, {Liu}, {L{\'o}pez-Coto}, {Lypova}, {Marandon},
  {Malyshev}, {Marcowith}, {Mariaud}, {Marx}, {Maurin}, {Maxted}, {Mayer},
  {Meintjes}, {Meyer}, {Mitchell}, {Moderski}, {Mohamed}, {Mohrmann},
  {Mor{\r{a}}}, {Moulin}, {Murach}, {Nakashima}, {de Naurois}, {Ndiyavala},
  {Niederwanger}, {Niemiec}, {Oakes}, {O'Brien}, {Odaka}, {Ohm}, {Ostrowski},
  {Oya}, {Padovani}, {Panter}, {Parsons}, {Paz Arribas}, {Pekeur}, {Pelletier},
  {Perennes}, {Petrucci}, {Peyaud}, {Piel}, {Pita}, {Poireau}, {Poon},
  {Prokhorov}, {Prokoph}, {P{\"u}hlhofer}, {Punch}, {Quirrenbach}, {Raab},
  {Rauth}, {Reimer}, {Reimer}, {Renaud}, {de los Reyes}, {Rieger}, {Rinchiuso},
  {Romoli}, {Rowell}, {Rudak}, {Rulten}, {Safi-Harb}, {Sahakian}, {Saito},
  {Sanchez}, {Santangelo}, {Sasaki}, {Schandri}, {Schlickeiser},
  {Sch{\"u}ssler}, {Schulz}, {Schwanke}, {Schwemmer}, {Seglar-Arroyo},
  {Settimo}, {Seyffert}, {Shafi}, {Shilon}, {Shiningayamwe}, {Simoni}, {Sol},
  {Spanier}, {Spir-Jacob}, {Stawarz}, {Steenkamp}, {Stegmann}, {Steppa},
  {Sushch}, {Takahashi}, {Tavernet}, {Tavernier}, {Taylor}, {Terrier},
  {Tibaldo}, {Tiziani}, {Tluczykont}, {Trichard}, {Tsirou}, {Tsuji}, {Tuffs},
  {Uchiyama}, {van der Walt}, {van Eldik}, {van Rensburg}, {van Soelen},
  {Vasileiadis}, {Veh}, {Venter}, {Viana}, {Vincent}, {Vink}, {Voisin},
  {V{\"o}lk}, {Vuillaume}, {Wadiasingh}, {Wagner}, {Wagner}, {Wagner}, {White},
  {Wierzcholska}, {Willmann}, {W{\"o}rnlein}, {Wouters}, {Yang}, {Zaborov},
  {Zacharias}, {Zanin}, {Zdziarski}, {Zech}, {Zefi}, {Ziegler}, {Zorn}, \&
  {{\.Z}ywucka}}]{hess18}
{H.~E.~S.~S. Collaboration}, {Abdalla}, H., {Abramowski}, A., {et~al.}
  2018{\natexlab{a}}, \aap, 612, A1, \dodoi{10.1051/0004-6361/201732098}

\bibitem[{{H.~E.~S.~S. Collaboration} {et~al.}(2018{\natexlab{b}}){H.~E.~S.~S.
  Collaboration}, {Abdalla}, {Abramowski}, {Aharonian}, {Ait Benkhali},
  {Akhperjanian}, {Andersson}, {Ang{\"u}ner}, {Arrieta}, {Aubert}, {Backes},
  {Balzer}, {Barnard}, {Becherini}, {Becker Tjus}, {Berge}, {Bernhard},
  {Bernl{\"o}hr}, {Blackwell}, {B{\"o}ttcher}, {Boisson}, {Bolmont}, {Bordas},
  {Bregeon}, {Brun}, {Brun}, {Bryan}, {Bulik}, {Capasso}, {Carr}, {Carrigan},
  {Casanova}, {Cerruti}, {Chakraborty}, {Chalme-Calvet}, {Chaves}, {Chen},
  {Chevalier}, {Chr{\'e}tien}, {Colafrancesco}, {Cologna}, {Condon}, {Conrad},
  {Couturier}, {Cui}, {Davids}, {Degrange}, {Deil}, {Devin}, {deWilt},
  {Dirson}, {Djannati-Ata{\"\i}}, {Domainko}, {Donath}, {Drury}, {Dubus},
  {Dutson}, {Dyks}, {Edwards}, {Egberts}, {Eger}, {Ernenwein}, {Eschbach},
  {Farnier}, {Fegan}, {Fernandes}, {Fiasson}, {Fontaine}, {F{\"o}rster},
  {Funk}, {F{\"u}{\ss}ling}, {Gabici}, {Gajdus}, {Gallant}, {Garrigoux},
  {Giavitto}, {Giebels}, {Glicenstein}, {Gottschall}, {Goyal}, {Grondin},
  {Hadasch}, {Hahn}, {Haupt}, {Hawkes}, {Heinzelmann}, {Henri}, {Hermann},
  {Hervet}, {Hillert}, {Hinton}, {Hofmann}, {Hoischen}, {Holler}, {Horns},
  {Ivascenko}, {Jacholkowska}, {Jamrozy}, {Janiak}, {Jankowsky}, {Jankowsky},
  {Jingo}, {Jogler}, {Jouvin}, {Jung-Richardt}, {Kastendieck},
  {Katarzy{\'n}ski}, {Katz}, {Kerszberg}, {Kh{\'e}lifi}, {Kieffer}, {King},
  {Klepser}, {Klochkov}, {Klu{\'z}niak}, {Kolitzus}, {Komin}, {Kosack},
  {Krakau}, {Kraus}, {Krayzel}, {Kr{\"u}ger}, {Laffon}, {Lamanna}, {Lau},
  {Lees}, {Lefaucheur}, {Lefranc}, {Lemi{\`e}re}, {Lemoine-Goumard}, {Lenain},
  {Leser}, {Lohse}, {Lorentz}, {Liu}, {L{\'o}pez-Coto}, {Lypova}, {Marandon},
  {Marcowith}, {Mariaud}, {Marx}, {Maurin}, {Maxted}, {Mayer}, {Meintjes},
  {Meyer}, {Mitchell}, {Moderski}, {Mohamed}, {Mohrmann}, {Mor{\r{a}}},
  {Moulin}, {Murach}, {de Naurois}, {Niederwanger}, {Niemiec}, {Oakes},
  {O'Brien}, {Odaka}, {{\"O}ttl}, {Ohm}, {de O{\~n}a Wilhelmi}, {Ostrowski},
  {Oya}, {Padovani}, {Panter}, {Parsons}, {Paz Arribas}, {Pekeur}, {Pelletier},
  {Perennes}, {Petrucci}, {Peyaud}, {Pita}, {Poon}, {Prokhorov}, {Prokoph},
  {P{\"u}hlhofer}, {Punch}, {Quirrenbach}, {Raab}, {Reimer}, {Reimer},
  {Renaud}, {de los Reyes}, {Rieger}, {Romoli}, {Rosier-Lees}, {Rowell},
  {Rudak}, {Rulten}, {Sahakian}, {Salek}, {Sanchez}, {Santangelo}, {Sasaki},
  {Schlickeiser}, {Sch{\"u}ssler}, {Schulz}, {Schwanke}, {Schwemmer},
  {Settimo}, {Seyffert}, {Shafi}, {Shilon}, {Simoni}, {Sol}, {Spanier},
  {Spengler}, {Spies}, {Stawarz}, {Steenkamp}, {Stegmann}, {Stinzing}, {Stycz},
  {Sushch}, {Tavernet}, {Tavernier}, {Taylor}, {Terrier}, {Tibaldo}, {Tiziani},
  {Tluczykont}, {Trichard}, {Tuffs}, {Uchiyama}, {Valerius}, {van der Walt},
  {van Eldik}, {van Soelen}, {Vasileiadis}, {Veh}, {Venter}, {Viana},
  {Vincent}, {Vink}, {Voisin}, {V{\"o}lk}, {Vuillaume}, {Wadiasingh}, {Wagner},
  {Wagner}, {Wagner}, {White}, {Wierzcholska}, {Willmann}, {W{\"o}rnlein},
  {Wouters}, {Yang}, {Zabalza}, {Zaborov}, {Zacharias}, {Zdziarski}, {Zech},
  {Zefi}, {Ziegler}, \& {{\.Z}ywucka}}]{hesspwn18}
---. 2018{\natexlab{b}}, \aap, 612, A2, \dodoi{10.1051/0004-6361/201629377}

\bibitem[{{H.~E.~S.~S. Collaboration} {et~al.}(2018{\natexlab{c}}){H.~E.~S.~S.
  Collaboration}, {Abdalla}, {Aharonian}, {Ait Benkhali}, {Ang{\"u}ner},
  {Arakawa}, {Arcaro}, {Armand}, {Arrieta}, {Backes}, {Barnard}, {Becherini},
  {Becker Tjus}, {Berge}, {Bernhard}, {Bernl{\"o}hr}, {Blackwell},
  {B{\"o}ttcher}, {Boisson}, {Bolmont}, {Bonnefoy}, {Bordas}, {Bregeon},
  {Brun}, {Brun}, {Bryan}, {B{\"u}chele}, {Bulik}, {Bylund}, {Capasso},
  {Caroff}, {Carosi}, {Casanova}, {Cerruti}, {Chakraborty}, {Chandra},
  {Chaves}, {Chen}, {Colafrancesco}, {Condon}, {Davids}, {Deil}, {Devin},
  {deWilt}, {Dirson}, {Djannati-Ata{\"\i}}, {Dmytriiev}, {Donath},
  {Doroshenko}, {Drury}, {Dyks}, {Egberts}, {Emery}, {Ernenwein}, {Eschbach},
  {Fegan}, {Fiasson}, {Fontaine}, {Funk}, {F{\"u}{\ss}ling}, {Gabici},
  {Gallant}, {Gat{\'e}}, {Giavitto}, {Glawion}, {Glicenstein}, {Gottschall},
  {Grondin}, {Hahn}, {Haupt}, {Heinzelmann}, {Henri}, {Hermann}, {Hinton},
  {Hofmann}, {Hoischen}, {Holch}, {Holler}, {Horns}, {Huber}, {Iwasaki},
  {Jacholkowska}, {Jamrozy}, {Jankowsky}, {Jankowsky}, {Jouvin},
  {Jung-Richardt}, {Kastendieck}, {Katarzy{\'n}ski}, {Katsuragawa}, {Katz},
  {Kerszberg}, {Khangulyan}, {Kh{\'e}lifi}, {King}, {Klepser}, {Klu{\'z}niak},
  {Komin}, {Kosack}, {Krakau}, {Kraus}, {Kr{\"u}ger}, {Lamanna}, {Lau},
  {Lefaucheur}, {Lemi{\`e}re}, {Lemoine-Goumard}, {Lenain}, {Leser}, {Lohse},
  {Lorentz}, {L{\'o}pez-Coto}, {Lypova}, {Malyshev}, {Marandon}, {Marcowith},
  {Mariaud}, {Mart{\'\i}-Devesa}, {Marx}, {Maurin}, {Meintjes}, {Mitchell},
  {Moderski}, {Mohamed}, {Mohrmann}, {Moulin}, {Murach}, {Nakashima}, {de
  Naurois}, {Ndiyavala}, {Niederwanger}, {Niemiec}, {Oakes}, {O'Brien},
  {Odaka}, {Ohm}, {Ostrowski}, {Oya}, {Padovani}, {Panter}, {Parsons},
  {Perennes}, {Petrucci}, {Peyaud}, {Piel}, {Pita}, {Poireau}, {Priyana Noel},
  {Prokhorov}, {Prokoph}, {P{\"u}hlhofer}, {Punch}, {Quirrenbach}, {Raab},
  {Rauth}, {Reimer}, {Reimer}, {Renaud}, {Rieger}, {Rinchiuso}, {Romoli},
  {Rowell}, {Rudak}, {Ruiz-Velasco}, {Sahakian}, {Saito}, {Sanchez},
  {Santangelo}, {Sasaki}, {Schlickeiser}, {Sch{\"u}ssler}, {Schulz},
  {Schwanke}, {Schwemmer}, {Seglar-Arroyo}, {Senniappan}, {Seyffert}, {Shafi},
  {Shilon}, {Shiningayamwe}, {Simoni}, {Sinha}, {Sol}, {Spanier}, {Specovius},
  {Spir-Jacob}, {Stawarz}, {Steenkamp}, {Stegmann}, {Steppa}, {Takahashi},
  {Tavernet}, {Tavernier}, {Taylor}, {Terrier}, {Tibaldo}, {Tiziani},
  {Tluczykont}, {Trichard}, {Tsirou}, {Tsuji}, {Tuffs}, {Uchiyama}, {van der
  Walt}, {van Eldik}, {van Rensburg}, {van Soelen}, {Vasileiadis}, {Veh},
  {Venter}, {Vincent}, {Vink}, {Voisin}, {V{\"o}lk}, {Vuillaume}, {Wadiasingh},
  {Wagner}, {Wagner}, {White}, {Wierzcholska}, {Yang}, {Zaborov}, {Zacharias},
  {Zanin}, {Zdziarski}, {Zech}, {Zefi}, {Ziegler}, {Zorn}, {{\.Z}ywucka},
  {Kerr}, {Johnston}, \& {Shannon}}]{hvela18}
{H.~E.~S.~S. Collaboration}, {Abdalla}, H., {Aharonian}, F., {et~al.}
  2018{\natexlab{c}}, \aap, 620, A66, \dodoi{10.1051/0004-6361/201732153}

\bibitem[{{HI4PI Collaboration} {et~al.}(2016){HI4PI Collaboration}, {Ben
  Bekhti}, {Fl{\"o}er}, {Keller}, {Kerp}, {Lenz}, {Winkel}, {Bailin},
  {Calabretta}, {Dedes}, {Ford}, {Gibson}, {Haud}, {Janowiecki}, {Kalberla},
  {Lockman}, {McClure-Griffiths}, {Murphy}, {Nakanishi}, {Pisano}, \&
  {Staveley-Smith}}]{nh16}
{HI4PI Collaboration}, {Ben Bekhti}, N., {Fl{\"o}er}, L., {et~al.} 2016, \aap,
  594, A116, \dodoi{10.1051/0004-6361/201629178}

\bibitem[{{Hobbs} {et~al.}(2006){Hobbs}, {Edwards}, \& {Manchester}}]{hob2006}
{Hobbs}, G.~B., {Edwards}, R.~T., \& {Manchester}, R.~N. 2006, \mnras, 369,
  655, \dodoi{10.1111/j.1365-2966.2006.10302.x}

\bibitem[{{Kothes} {et~al.}(2006){Kothes}, {Fedotov}, {Foster}, \&
  {Uyan{\i}ker}}]{kot+06}
{Kothes}, R., {Fedotov}, K., {Foster}, T.~J., \& {Uyan{\i}ker}, B. 2006, \aap,
  457, 1081, \dodoi{10.1051/0004-6361:20065062}

\bibitem[{{Kothes} {et~al.}(2008){Kothes}, {Landecker}, {Reich}, {Safi-Harb},
  \& {Arzoumanian}}]{kot+08}
{Kothes}, R., {Landecker}, T.~L., {Reich}, W., {Safi-Harb}, S., \&
  {Arzoumanian}, Z. 2008, \apj, 687, 516, \dodoi{10.1086/591653}

\bibitem[{{Landecker} {et~al.}(1990){Landecker}, {Clutton-Brock}, \&
  {Purton}}]{lcp90}
{Landecker}, T.~L., {Clutton-Brock}, M., \& {Purton}, C.~R. 1990, \aap, 232,
  207

\bibitem[{{Linden} {et~al.}(2017){Linden}, {Auchettl}, {Bramante}, {Cholis},
  {Fang}, {Hooper}, {Karwal}, \& {Li}}]{lin+17}
{Linden}, T., {Auchettl}, K., {Bramante}, J., {et~al.} 2017, \prd, 96, 103016,
  \dodoi{10.1103/PhysRevD.96.103016}

\bibitem[{{Nolan} {et~al.}(2012){Nolan}, {Abdo}, {Ackermann}, {Ajello},
  {Allafort}, {Antolini}, {Atwood}, {Axelsson}, {Baldini}, {Ballet}, \&
  et~al.}]{nol+12}
{Nolan}, P.~L., {Abdo}, A.~A., {Ackermann}, M., {et~al.} 2012, \apjs, 199, 31,
  \dodoi{10.1088/0067-0049/199/2/31}

\bibitem[{{Pivato} {et~al.}(2013){Pivato}, {Hewitt}, {Tibaldo}, {Acero},
  {Ballet}, {Brandt}, {de Palma}, {Giordano}, {Janssen}, {J{\'o}hannesson}, \&
  {Smith}}]{piv+13}
{Pivato}, G., {Hewitt}, J.~W., {Tibaldo}, L., {et~al.} 2013, \apj, 779, 179,
  \dodoi{10.1088/0004-637X/779/2/179}

\bibitem[{{Porter} {et~al.}(2006){Porter}, {Moskalenko}, \&
  {Strong}}]{Porter2006}
{Porter}, T.~A., {Moskalenko}, I.~V., \& {Strong}, A.~W. 2006, \apjl, 648, L29,
  \dodoi{10.1086/507770}

\bibitem[{{Ray} {et~al.}(2011){Ray}, {Kerr}, {Parent}, {Abdo}, {Guillemot},
  {Ransom}, {Rea}, {Wolff}, {Makeev}, {Roberts}, {Camilo}, {Dormody}, {Freire},
  {Grove}, {Gwon}, {Harding}, {Johnston}, {Keith}, {Kramer}, {Michelson},
  {Romani}, {Saz Parkinson}, {Thompson}, {Weltevrede}, {Wood}, \&
  {Ziegler}}]{ray2011}
{Ray}, P.~S., {Kerr}, M., {Parent}, D., {et~al.} 2011, \apjs, 194, 17,
  \dodoi{10.1088/0067-0049/194/2/17}

\bibitem[{{Saz Parkinson} {et~al.}(2010){Saz Parkinson}, {Dormody}, {Ziegler},
  {Ray}, {Abdo}, {Ballet}, {Baring}, {Belfiore}, {Burnett}, {Caliandro},
  {Camilo}, {Caraveo}, {de Luca}, {Ferrara}, {Freire}, {Grove}, {Gwon},
  {Harding}, {Johnson}, {Johnson}, {Johnston}, {Keith}, {Kerr},
  {Kn{\"o}dlseder}, {Makeev}, {Marelli}, {Michelson}, {Parent}, {Ransom},
  {Reimer}, {Romani}, {Smith}, {Thompson}, {Watters}, {Weltevrede}, {Wolff}, \&
  {Wood}}]{par+10}
{Saz Parkinson}, P.~M., {Dormody}, M., {Ziegler}, M., {et~al.} 2010, \apj, 725,
  571, \dodoi{10.1088/0004-637X/725/1/571}

\bibitem[{{Shibata} {et~al.}(2011){Shibata}, {Ishikawa}, \&
  {Sekiguchi}}]{shibata11}
{Shibata}, T., {Ishikawa}, T., \& {Sekiguchi}, S. 2011, \apj, 727, 38,
  \dodoi{10.1088/0004-637X/727/1/38}

\bibitem[{{Tian} \& {Leahy}(2006)}]{tl06}
{Tian}, W.~W., \& {Leahy}, D.~A. 2006, \aap, 455, 1053,
  \dodoi{10.1051/0004-6361:20065140}

\bibitem[{{VERITAS Collaboration} {et~al.}(2011){VERITAS Collaboration},
  {Aliu}, {Arlen}, {Aune}, {Beilicke}, {Benbow}, {Bouvier}, {Bradbury},
  {Buckley}, {Bugaev}, {Byrum}, {Cannon}, {Cesarini}, {Christiansen}, {Ciupik},
  {Collins-Hughes}, {Connolly}, {Cui}, {Dickherber}, {Duke}, {Errando},
  {Falcone}, {Finley}, {Finnegan}, {Fortson}, {Furniss}, {Galante}, {Gall},
  {Gibbs}, {Gillanders}, {Godambe}, {Griffin}, {Grube}, {Guenette}, {Gyuk},
  {Hanna}, {Holder}, {Huan}, {Hughes}, {Hui}, {Humensky}, {Imran}, {Kaaret},
  {Karlsson}, {Kertzman}, {Kieda}, {Krawczynski}, {Krennrich}, {Lang},
  {Lyutikov}, {Madhavan}, {Maier}, {Majumdar}, {McArthur}, {McCann},
  {McCutcheon}, {Moriarty}, {Mukherjee}, {Nu{\~n}ez}, {Ong}, {Orr}, {Otte},
  {Park}, {Perkins}, {Pizlo}, {Pohl}, {Prokoph}, {Quinn}, {Ragan}, {Reyes},
  {Reynolds}, {Roache}, {Rose}, {Ruppel}, {Saxon}, {Schroedter}, {Sembroski},
  {{\c{S}}ent{\"u}rk}, {Smith}, {Staszak}, {Te{\v{s}}i{\'c}}, {Theiling},
  {Thibadeau}, {Tsurusaki}, {Tyler}, {Varlotta}, {Vassiliev}, {Vincent},
  {Vivier}, {Wakely}, {Ward}, {Weekes}, {Weinstein}, {Weisgarber}, {Williams},
  \& {Zitzer}}]{vcrab11}
{VERITAS Collaboration}, {Aliu}, E., {Arlen}, T., {et~al.} 2011, Science, 334,
  69, \dodoi{10.1126/science.1208192}
  
\bibitem[Abeysekara et al.(2018)]{abe+18} Abeysekara, A.~U., Archer, A., Benbow, W., et al.\ 2018, \apj, 866, 24. doi:10.3847/1538-4357/aade4e

\bibitem[{{Xing} {et~al.}(2015){Xing}, {Wang}, {Zhang}, \& {Chen}}]{xing+15}
{Xing}, Y., {Wang}, Z., {Zhang}, X., \& {Chen}, Y. 2015, \apj, 805, 19,
  \dodoi{10.1088/0004-637X/805/1/19}

\bibitem[{{Zabalza}(2015)}]{naima}
{Zabalza}, V. 2015, Proc.~of International Cosmic Ray Conference 2015, 922

\bibitem[{{Zhang} \& {Liu}(2019)}]{zhang2019}
{Zhang}, X., \& {Liu}, S. 2019, \apj, 876, 24, \dodoi{10.3847/1538-4357/ab14df}

\bibitem[{{Zhu} {et~al.}(2018){Zhu}, {Zhang}, \& {Fang}}]{zzf18}
{Zhu}, B.-T., {Zhang}, L., \& {Fang}, J. 2018, \aap, 609, A110,
  \dodoi{10.1051/0004-6361/201629108}

\end{thebibliography}


\clearpage
\begin{table}
\begin{center}
\tabletypesize{\footnotesize}
\tablecolumns{10}
\tablewidth{240pt}
\setlength{\tabcolsep}{2pt}
\caption{Timing parameters for PSR~\psr.}
\label{tab:search_result}
\begin{tabular}{lccc}
\hline
	& $f$ & $f1/10^{-12}$ & $f2/10^{-22}$ \\
	& (Hz) & (Hz~s$^{-1}$) & (Hz~s$^{-2}$) \\ \hline
Known ephemeris & 10.7863425015(2) & -2.462252(6) & 0.122(7) \\
MJD 54682--57500 & 10.7863425017(4) & -2.46225(5) & 0.11(7) \\
MJD 57500--59470 & 10.786340(3) & -2.41(6) & -7(9) \\
\hline
\end{tabular}
\vskip 1mm
\footnotesize{Note: the frequency epoch is MJD 55225.}
\end{center}
\end{table}

\begin{table}
\begin{center}
\tabletypesize{\footnotesize}
\tablecolumns{10}
\tablewidth{240pt}
\setlength{\tabcolsep}{2pt}
\caption{Binned likelihood analysis results for the different spin phase data.}
\label{tab:likelihood}
\begin{tabular}{lccccccc}
\hline
	Phase range (src) & $F_{0.1-500}/10^{-8}$ & $\Gamma$ & $E_{c}$ & TS & $-2\log(L_{\rm ext}/L_{\rm src})$ \\
	&  (photons~s$^{-1}$\,cm$^{-2}$) & & (GeV) & &  \\ \hline
Onpulse ($-$) & 23.7$\pm$0.5 & 1.42$\pm$0.01 & 1.60$\pm$0.03 & 14030 & $-$ \\ \hline
Bridge ($-$) & 8.4$\pm$0.6 & 1.29$\pm$0.06 & 0.96$\pm$0.07 & 1664 & $-$ \\ \hline
Offpulse (PS1) & 1.8$\pm$0.6 & 2.4$\pm$0.1 & $-$ & 37& 16 \\ \hline
Offpulse (2PS, PS1) & 1.2$\pm$0.5 & 2.3$\pm$0.1 & $-$ & 26 & 23 \\
Offpulse (2PS, PS2) & 1.7$\pm$0.8 & 2.6$\pm$0.2 & $-$ & 18 &  \\ \hline
\hline
\end{tabular}
\vskip 1mm
\footnotesize{Note: $F_{0.1-500}$ is the photon flux in 0.1--500 GeV energy range.}
\end{center}
\end{table}

\begin{table}
\begin{center}
\tabletypesize{\footnotesize}
\tablecolumns{10}
\tablewidth{240pt}
\setlength{\tabcolsep}{2pt}
\caption{Spectral data points for PSR~\psr\ and PS1}
\label{tab:spectra}
\begin{tabular}{lccccccccccccc}
\hline
	$E$ & Band & $F/10^{-11}$ (Onpulse) & $F/10^{-11}$ (Bridge) & $F/10^{-11}$ (Offpulse-PS1) \\
	(GeV) & (GeV) & (erg cm$^{-2}$ s$^{-1}$) & (erg cm$^{-2}$ s$^{-1}$) &  (erg cm$^{-2}$ s$^{-1}$) \\ \hline
0.08 & 0.05--0.13 & 1.9$\pm$0.3 & 1.23 & 1.02 \\
0.20 & 0.13--0.32 & 3.8$\pm$0.3 & 1.6$\pm$0.3 & 0.69 \\
0.50 & 0.32--0.79 & 5.1$\pm$0.1 & 1.8$\pm$0.1 & 0.4$\pm$0.1 \\
1.26 & 0.79--1.99 & 5.9$\pm$0.1 & 2.1$\pm$0.1 & 0.15$\pm$0.06 \\
3.15 & 1.99--5.00 & 4.6$\pm$0.1 & 1.5$\pm$0.1 & 0.13$\pm$0.05 \\
7.92 & 5.00--12.56 & 2.4$\pm$0.1 & 0.42$\pm$0.08 & 0.13 \\
19.91 & 12.56--31.55 & 0.6$\pm$0.2 & 0.20 & 0.13 \\
50.00 & 31.55--79.24 & 0.18$\pm$0.09 & 0.10 & 0.30 \\
125.59 & 79.24--199.05 & 0.13 & 0.66 & 0.16 \\
315.48 & 199.05--500.0 & 0.2$\pm$0.1 & 0.47 & 0.40 \\
\hline
\end{tabular}
\vskip 1mm
\footnotesize{Note: $F$ is the energy flux ($E^{2}dN/dE$), and fluxes without uncertainties are the 95$\%$ upper limits.}
\end{center}
\end{table}
\end{document}